\documentclass[lettersize,journal]{IEEEtran}

\usepackage{tikz}

\usepackage[lined,ruled,commentsnumbered]{algorithm2e}

\usepackage{cite}
\usepackage[numbers,sort&compress]{}
\usepackage{amsmath,amssymb,amsfonts}
\usepackage{array}
\usepackage{graphicx}
\usepackage{makecell} 

\usepackage{textcomp}
\usepackage{xcolor}
\usepackage{booktabs}
\usepackage{multirow}
\usepackage{subfigure}
\usepackage{longtable}
\usepackage{threeparttable}
\usepackage{enumitem}
\usepackage{amssymb}
\usepackage{soul}
\usepackage{textcomp}
\usepackage{xcolor}
\usepackage{array}
\usepackage{url}
\usepackage{float}
\usepackage{bm}
\usepackage{adjustbox}
\usepackage{epstopdf}

\usepackage{diagbox}

\usepackage{amsthm}
\usepackage{tabularx}
\usepackage{mathrsfs} 
\usepackage{soul}
\usepackage{cuted}   
\usepackage{stfloats}

\theoremstyle{remark}  

\usepackage{algpseudocode}

\usepackage{stfloats}
\usepackage[colorlinks=true, linkcolor=blue, citecolor=blue]{hyperref}

\usepackage{cleveref}
\begin{document}
\newcommand{\minitab}[2][l]{\begin{tabular}{#1}#2\end{tabular}}

\def\BibTeX{{\rm B\kern-.05em{\sc i\kern-.025em b}\kern-.08em
    T\kern-.1667em\lower.7ex\hbox{E}\kern-.125emX}}


\title{AmbSentry: Mitigating Sensing Eavesdropping in ISAC Systems by Harnessing Ambient IoT Devices}

\author{Yifan~Zhang,~\IEEEmembership{Member,~IEEE,}
Yu~Bai,~\IEEEmembership{Member,~IEEE,}
        Riku Jäntti,~\IEEEmembership{Senior Member,~IEEE,}\\
        Zhu~Han,~\IEEEmembership{Fellow,~IEEE,}
        and Christos Masouros,~\IEEEmembership{Fellow,~IEEE.} 
\thanks{This manuscript is supported by the Smart Networks and Services Joint Undertaking under the European Union’s Horizon Europe research and innovation program (Ambient-6G) under Grant Agreement No 101192113.}
\thanks{Y. F. Zhang and R. Jäntti are with the Department of Information and Communications Engineering, Aalto University, Espoo, 02150, Finland. (email: yifan.1.zhang@aalto.fi, riku.jantti@aalto.fi)}
\thanks{Y. Bai is with the School of Software, Taiyuan University of Technology, Taiyuan 030024, China (e-mail: yu.baielec@gmail.com). }
\thanks{Z. Han is with the Department of Electrical and Computer Engineering, University of Houston, Houston, TX 77004 USA (email: hanzhu22@gmail.com)}
\thanks{C. Masouros is with the Department of Electronic and Electrical
Engineering, University College London, London WC1E 7JE, U.K. (e-mail:
chris.masouros@ieee.org).}
 }

\markboth{}%
{Shell \MakeLowercase{\textit{et al.}}: A Sample Article Using IEEEtran.cls for IEEE Journals}
\maketitle

\begin{abstract}
Integrated sensing and communication (ISAC) has emerged as a pivotal paradigm for 6G networks, enabling the synergistic convergence of spectral and hardware resources to maximize system efficiency. However, the inherent openness of wireless transmission exposes ISAC systems to critical security risks, particularly regarding the privacy of the sensing information. Unauthorized sensing eavesdroppers can extract sensitive target parameters (e.g., range and velocity) by directly estimating open sensing echo channels, rendering traditional data-based protection techniques ineffective.
To mitigate this threat, this paper proposes AmbSentry, an ISAC system that prevents the leakage of sensing information to sensing eavesdroppers by harnessing naturally distributed passive ambient IoT (AIoT) devices.
Specifically, these AIoT devices are strategically configured to act as cooperative jammers and ghost targets, introducing controllable interference into the sensing environment. Based on the proposed system, we formulate a joint optimization problem to maximize the integrated sidelobe level at the eavesdropper under quality-of-service (QoS) constraints, thereby degrading sensing eavesdropping performance while maintaining sensing and communication performance for legitimate receivers.
Since the problem is non-convex, we further develop an efficient iterative algorithm to cooperatively design the transmit beamforming at the base station and the reflection modulations of the AIoT devices based on Dinkelbach transformation and block coordinate descent methods. Finally, we conduct a comprehensive simulation to reveal the fundamental tradeoffs between sensing security and the QoS performance in AmbSentry. The detailed results also demonstrate that AmbSentry significantly enhances sensing security, allowing the legitimate sensing receiver to achieve a 14-dB SNR advantage in detection probability and a hundred times lower estimation error compared to the eavesdropper.

\end{abstract}


\section{Introduction} \label{s1_intro}

Integrated sensing and communication (ISAC) is a promising technology for 6G and beyond wireless systems~\cite{zhu2024enabling,liu2022integrated}, driven by its capability to simultaneously provide communication and radar sensing functionalities. 
However, integrating sensing capabilities into communication networks introduces unique security challenges, particularly the potential leakage of sensing information.
 The inherent openness of the wireless medium allows unauthorized devices to leverage intercepted transmit waveforms and the corresponding reflected echoes to reconstruct sensing-related observations, enabling unauthorized tracking, localization, or identification~\cite{10587082, chen2025sensing}.
Meanwhile, conventional data-based security methods provide limited protection in this situation, since sensing information is not embedded in the digital payload but is inherently encoded in the physical-layer propagation signatures of the waveform (e.g., delay and Doppler shift). 

To prevent communication and sensing eavesdropping in ISAC systems, various physical layer security (PLS) schemes have been proposed, leveraging the inherent properties of wireless propagation (e.g., noise,
interference, and fading) to protect the transmitted waveform.
To prevent communication eavesdropping, artificial noise (AN) injection~\cite{10227884,10781436,11016735} and auxiliary devices such as reconfigurable intelligent surfaces (RIS) and unmanned aerial vehicles (UAVs)~\cite{10870062,10616025,10238433} 
have been employed. These methods effectively jam eavesdroppers by generating AN while maintaining legitimate communication performance. However, they often incur significant energy overhead or high deployment costs, thus constraining their practical scalability. Meanwhile, the challenge of preventing sensing eavesdroppers remains largely unaddressed. Existing studies have attempted to mitigate this by optimizing detection probabilities~\cite{10605793}, minimizing the Cramér-Rao bound (CRB)~\cite{10804654}, and using ambiguity function engineering that designs transmitting signal to generate false targets~\cite{11202391}. Nevertheless, these approaches face some limitations, including the unrealistic assumption of knowing the channel state information (CSI) of a passive eavesdropper~\cite{10804654} or the resulting performance degradation for legitimate receivers~\cite{11202391}. 
Consequently, there is a growing need to develop an effective security paradigm that resists sensing eavesdroppers without their CSI or location information while maintaining effective quality of service (QoS).

Ambient IoT (AIoT) devices, such as RFID tags, passive backscatter sensors, and smart labels, have recently attracted considerable interest due to their widespread deployment~\cite{10616210, famaey2026survey}.
By passively reflecting and modulating incident radio frequency (RF) signals, AIoT devices eliminate the need for power-hungry RF chains \cite{ren2023toward}, thereby achieving ultra-low power consumption and low hardware complexity.
In addition, their compact and simple design makes them extremely cost-effective, with unit costs of 7 to 15 cents (USD)~\cite{8368232}, and a small size within a few square centimeters~\cite{abdulghafor2021recent}. These distinctive features have led to the widespread adoption of AIoT devices and their incorporation into the 3rd Generation Partnership Project (3GPP) standardization~\cite{3gppTR22840}. By strategically modulating the reflected signal, AIoT devices can introduce controllable randomness and multiplicative noise into the propagation environment~\cite{11500508}. This stochastic interference can effectively obfuscate eavesdroppers' sensing observations, thereby preventing unauthorized reconstruction of sensing privacy. Consequently, AIoT device-aided ISAC systems offer a potential solution to enhance ubiquitous sensing security in an energy-efficient and low-cost manner, bypassing the substantial power and cost penalties inherent in conventional eavesdropper-jamming methods.

\subsection{Related Works}

\subsubsection{Preventing Communication Eavesdropping in ISAC systems} The most recent literature on PLS in ISAC systems focuses on preventing communication eavesdropping activities. For example, sensing-assisted protocols have been proposed, where the location~\cite{10227884} or CSI~\cite{10781436} of potential eavesdroppers is first estimated at a base station (BS) and then exploited to allow the legitimate transmitter to precisely steer AN to jam the eavesdroppers. Further extensions have incorporated non-orthogonal multiple access (NOMA) to manage the trade-off between data rate and the security~\cite{11016735}, achieving a mutual performance gain for multi-user scenarios. However, reliance on AN to confuse eavesdroppers incurs a significant power penalty. This additional energy consumption diverts scarce resources away from the core sensing and communication (S\&C) tasks, inevitably degrading the overall energy efficiency of the ISAC system.

In addition, auxiliary devices have been integrated into ISAC security to circumvent line-of-sight obstructions and expand the spatial degrees of freedom (DoF) to enhance communication security. Specifically, RIS has been deployed to achieve ISAC security by specifically improving the secrecy rate for legitimate users~\cite{10870062}. Furthermore, a dynamic optimization of secure transmission strategies has been proposed in a UAV-mounted RIS system for data transmission security~\cite{10616025}. More recently, advanced hardware such as intelligent omni-surfaces (IOS)~\cite{10238433}, which supports simultaneous reflection and transmission, has been proposed to achieve full-space security. Despite their potential, the large-scale implementation of these devices faces practical barriers due to high deployment, maintenance, and control costs. Furthermore, while these auxiliary device-aided schemes have been extensively studied for communication security, their application to sensing security in ISAC systems has been largely overlooked.

\subsubsection{Preventing Sensing Eavesdropping   in ISAC systems} Recently, sensing-secure ISAC designs have been explored in~\cite{10605793,10804654,10587082}. These studies aim to prevent sensing eavesdropping by optimizing the probability of detection in ISAC systems~\cite{10605793} or minimizing the CRB for legitimate sensing~\cite{10804654}, while simultaneously suppressing the detection capabilities of eavesdroppers. Furthermore, signaling designs that incorporate mutual radar information have been proposed to maximize legitimate sensing performance~\cite{10587082}, subject to strict constraints on eavesdropper sensing performance and communication QoS. However, a fundamental limitation shared by these early studies is their reliance on the CSI of eavesdropping channels available to the legitimate ISAC transmitter. This assumption could be unrealistic, as a passive radar eavesdropper remains silent and does not reveal its location. To overcome this issue, an ambiguity function (AF) engineering method~\cite{11202391} has been proposed to shape the waveform autocorrelation, thereby creating artificial false targets in the transmitted waveform and effectively obfuscating the eavesdroppers' detection of the genuine target. Although legitimate receivers can recover the true target using a reciprocal filter and a priori AF knowledge, this waveform-shaping strategy inevitably compromises legitimate S\&C performance and requires complex receiver-side processing. Most recently, a sensing security method was proposed in~\cite{chen2025sensing} that illuminates other scatterers and clutter to confuse sensing eavesdroppers in a near-field ISAC system. However, this scheme specified the scenario and required the presence of known scatterers or clutter.

 \subsection{Motivations and Contributions}

Despite recent progress in ISAC security, existing solutions face {\em three fundamental challenges} that hinder their practical implementation.
First, active defense mechanisms, such as AN injection, incur additional energy overhead in AN transmission~\cite{zhang2025artificial}. 
Second, while hardware-assisted schemes (e.g., RIS, UAVs) offer performance gains, they entail substantial deployment and maintenance costs~\cite{zhang2025backscatter}, limiting their large-scale scalability.
Third, and most critically, current research predominantly focuses on communication secrecy or relies on the impractical assumption that the passive eavesdropper's CSI is known to achieve sensing security. 
Consequently, there is an urgent need to explore an effective sensing security method that can enhance sensing security without relying on knowledge of eavesdropper capabilities or excessive energy and cost consumption.



To this end, this paper proposes AmbSentry, a novel ISAC system that enhances sensing security by harnessing AIoT devices to prevent eavesdropping on sensing activities. Specifically, the passive nature of AIoT devices is leveraged to reflect incident ISAC signals in a specialized manner, thereby introducing controllable multiplicative noise and artificial clutter into the propagation environment. By exploiting its knowledge of the location and modulation of AIoT devices, a legitimate sensing receiver can suppress interference, whereas a passive sensing eavesdropper lacks this side information and thus experiences largely unmitigated artificial clutter, thereby establishing a practical sensing-performance gap.
To theoretically characterize system performance without eavesdropping CSI, we derive a closed-form expression for the integrated sidelobe level (ISL) as a metric for quantifying the jamming efficiency of AIoT devices against sensing eavesdroppers. Furthermore, we develop a joint signaling optimization strategy for transmitting beamforming and AIoT device modulation to maximize ISL performance. Finally, we present comprehensive simulation results, including ISL analysis, detection probability, and estimation performance, to validate the effectiveness of AmbSentry. Specifically, the primary contributions of this paper are summarized as follows:

 \begin{itemize} \item \textbf{Proposal of a Practical AIoT device-assisted ISAC System for Sensing Security.} We propose a practical AIoT device-assisted ISAC system that utilizes off-the-shelf AIoT devices naturally distributed in real-world environments for preventing sensing eavesdropping without requiring energy-intensive active jamming. 
Specifically, we strategically design these AIoT devices to serve as cooperative jammers that introduce controllable clutter to obfuscate eavesdroppers' sensing capabilities, while the
legitimate sensing receiver eliminates the interference using
prior knowledge of the AIoT devices.

\begin{figure}[tp]
    \centering
    \includegraphics[width=0.95\linewidth]{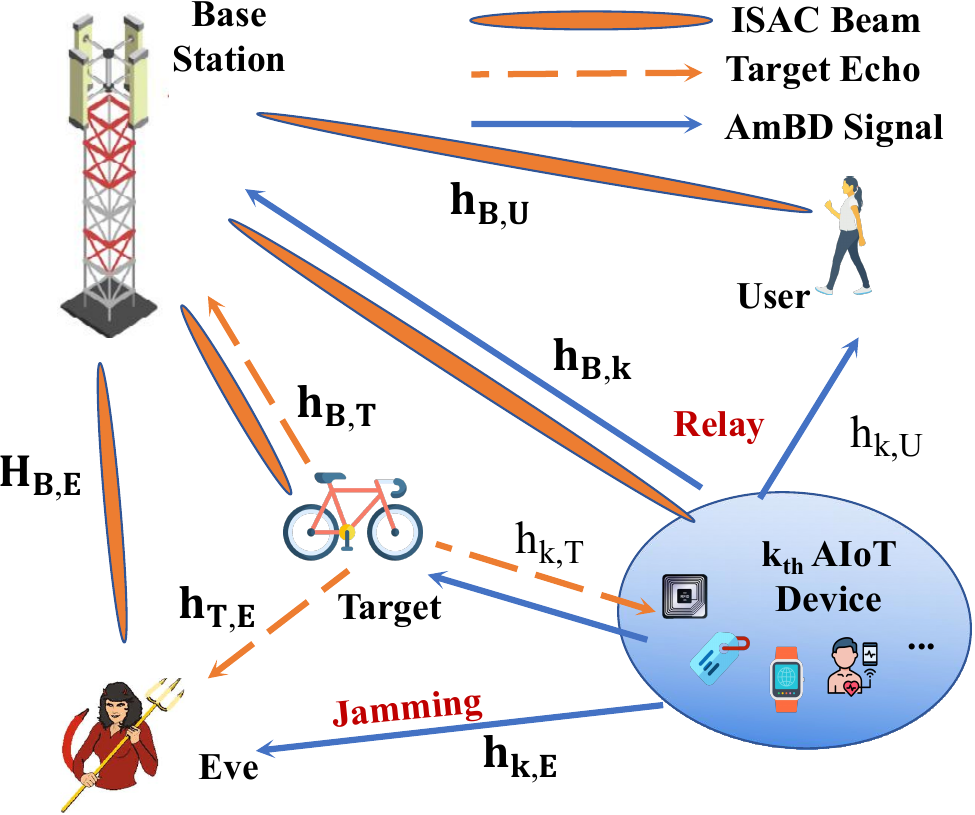}
    \caption{The system model of AmbSentry, where AIoT devices are leveraged as
friendly jammers that randomly reflect signals to jam eavesdroppers. The transmit signal from the BS and the backscattered signal from the AIoT device are jointly optimized for enhanced sensing security.}
    \label{syse}
\end{figure}

\item \textbf{Theoretical Analysis for Sensing Secrecy.} We establish a theoretical framework to quantify sensing security by deriving the closed-form expression for the ISL without needing the eavesdropping CSI. Subsequently, we formulate a sensing-secure problem that maximizes the ISL to impair the eavesdroppers' sensing capability, jointly optimizing the BS's transmit signal beamforming and the AIoT devices' backscatter signals.
Moreover, we consider the signal-to-interference ratio (SINR) and communication rate constraints to maintain the legitimate QoS of S\&C. The problem maximizes interference to eavesdroppers while maintaining the S\&C performance for legitimate receivers.

\item \textbf{Joint Transmit Beamforming and AIoT device Modulation Optimization.} 
We propose an iterative algorithm to jointly optimize the transmit beamforming vectors and AIoT device reflection coefficients. Specifically, the generalized Dinkelbach transformation and the block coordinate descent (BCD) method are first used to decompose the original problem into two subproblems: a transmit beamforming optimization subproblem and an AIoT device backscatter optimization subproblem. Then, we apply the successive convex approximation (SCA) semidefinite relaxation (SDR) methods to solve the non-convex subproblem.

\end{itemize}

Furthermore, extensive simulation results validate the effectiveness of AmbSentry in achieving robust sensing security. Specifically, the trade-offs between the sensing security and S\&C performance 
are first investigated, revealing how the transmit beamforming at the AIoT devices and the backscatter modulations of the AIoT devices influence ISAC system performance. Then, the detection and estimation performance for legitimate sensing receivers and eavesdroppers is evaluated.
The numerical results demonstrate enhanced security, achieving a performance gap of approximately 14~dB in detection probability and a significant range estimation gap between the eavesdropper and the legitimate sensing receiver, with the RMSE varying by three orders of magnitude, from $10^{-2}$~m for the legitimate sensing receiver to $10^1$~m for the eavesdropper.

\textit{Notations:} Boldface lower-case and upper-case letters represent vectors and matrices, respectively. 
The operators \((\cdot)^{{T}}\), \((\cdot)^{{H}}\), and \((\cdot)^{*}\) denote the transpose, the Hermitian transpose, and the complex conjugate, respectively.
$||\cdot||$ and $|\cdot|$ denote the Euclidean norm of a vector and the absolute value of a complex scalar, respectively. 
$\mathbb{E}\{\cdot\}$ stands for the statistical expectation operator. 
$\mathbb{C}^{M \times N}$ represents the space of complex matrices $M \times N$. 
$x \sim \mathcal{CN}(\mu, \sigma^2)$ indicates that the random variable $x$ follows a circularly symmetric complex Gaussian (CSCG) distribution with mean $\mu$ and variance $\sigma^2$. 
$\text{diag}(\cdot)$ denotes a diagonal matrix formed by the input vector or elements. 
$I_M$ represents the identity matrix of size $M \times M$. The trace of a square matrix, denoted by $\mathrm{tr}(\cdot)$, is defined as the sum of the elements on its main diagonal.



\section{System Model} \label{System model}

AmbSentry harnesses existing AIoT devices to establish a secure ISAC environment, aiming to obscure the true targets from an unknown sensing eavesdropper (SEve) that attempts to reconstruct target parameters from the physical waveforms. 
In contrast, communication-related eavesdropping has been extensively studied~\cite{10227884,10781436,11016735,10870062,10616025,10238433} and is thus beyond the scope of this work. Fig.~\ref{syse} illustrates a typical monostatic ISAC system~\cite{liu2022integrated}, in which a multi-antenna BS simultaneously communicates with a single-antenna user and senses multiple targets.
In addition, a set of naturally present spatially distributed single-antenna AIoT devices is leveraged, with their reflections configured as controllable interference to jam SEve.
The target and AIoT devices can be mobile or static. Additionally, a multiple-input-multiple-output (MIMO) system~\cite{9778563} is considered where the BS is equipped with $M_t$ transmit antennas and $M_r$ receive antennas, while SEve is equipped with $M_e$ receive antennas.

\subsection{Transmit Signal and Channel Models}

The transmitted ISAC signal, utilizing \(N\) orthogonal frequency division multiplexing (OFDM) subcarriers, is modulated with random communication symbols drawn from the constellation set \(\mathcal{S}\)~\cite{11037613}. Define the transmit signal $\mathbf{s} = [s[1], s[2], \ldots, s[n]]^{T}$,
where \(s[n] \in \mathcal{S}\) for all \(n = 0,2,\ldots,N-1\). Without loss of generality, the constellation symbols are assumed to be zero-mean and unit-variance, i.e., $\mathbb{E}\{s[n]\}=0, \mathbb{E}\{|s[n]|^2\}=1,\ \forall n$. Let \(\mathbf{w}\in\mathbb{C}^{M_t\times 1}\) denote the unit-norm transmit beamforming vector with \(\|\mathbf{w}\|^2 = 1\). The vector signal transmitted in the subcarrier \(n\) is
\begin{equation}
    \mathbf{x}[n] = \sqrt{P_t} \mathbf{w}  s[n],
\end{equation}
where \(P_t\) is the transmit power of the BS. Let $\mu_{2}=\mathbb{E}\{|s[n]|^{2}\}$ and $\mu_{4}=\mathbb{E}\{|s[n]|^{4}\}$ denote the second and fourth moments of the constellation~\cite{11087656}, respectively.

An OFDM channel model is adopted for all links in the considered monostatic ISAC system. The OFDM carrier frequencies are
\begin{equation}
f_n = f_c + (n-1)\Delta f,  n=0,1,\ldots,N-1,
\end{equation}
where \(f_c\) and \(\Delta f\) denote carrier frequency and subcarrier spacing, respectively. 
Without loss of generality, for any link from a transmitter node \(q\) with \(M_q\) antennas to a receiver node \(p\) with \(M_p\) antennas, the frequency-domain MIMO channel on subcarrier \(n\) is denoted as
\(\mathbf{h}_{q p}[n]\in\mathbb{C}^{M_q\times M_p}\), and is modeled as~\cite{6847111}
\begin{equation}\label{eq:geom_channel}
\mathbf{h}_{qp}[n]
= \sqrt{\beta_{q p}}
\sum_{\ell=0}^{L_{q p}-1}
\tilde g_{q p,\ell} 
\mathbf{a}_{q}(\theta_{q p,\ell})
 e^{-j2\pi f_n \tau_{q p,\ell}},
\end{equation}
where \(L_{q p}\) is the number of multipath components, 
\(\beta_{q p}\) captures the large-scale attenuation, 
\(\tilde g_{q p,\ell}\) denotes the effective complex path gain that incorporates the small-scale fading and the transmit-side response, 
\(\tau_{q p,\ell}\) is the path delay, and 
\(\mathbf{a}_{q}(\theta_{q p,\ell})\) is the receive array response associated with the angle of arrival \(\theta_{q p,\ell}\).

\subsection{Backscatter Model}

Consider a set of $K$ single-antenna AIoT devices, indexed by $k \in \mathcal{K} = \{1, \dots, K\}$. Let $\mathbf{h}_{B,k} \in \mathbb{C}^{1 \times M_t}$ denote the channel vector from the BS to the $k$-th AIoT device. Similarly, the channel response from the $k$-th AIoT device to the user is represented by $h_{k,U} \in \mathbb{C}$ and the channel from the $k$-th AIoT device to SEve is denoted as $\mathbf{h}_{k,E} \in \mathbb{C}^{M_e \times 1}$. Consequently, the signal transmitted by the BS impinging on the $k$-th AIoT device at the $n$-th subcarrier is given by 
\begin{equation}
    r_k[n]
    =  \mathbf{h}_{B,k}\mathbf{x}[n]
    =  \sqrt{P_t}\mathbf{h}_{B,k}\mathbf{w}  s[n] .
\end{equation}

The \(\mathrm{ k_{th}}\) AIoT device modulates the incident signal by adjusting its amplitude reflection coefficient $\alpha_k$. Considering practical $L$-ary ASK modulation of AIoT devices~\cite{8809276}, $\alpha_k$ is selected from a finite discrete set $\mathcal{A}$, where $\mathcal{A} = \{\alpha_{1},\alpha_{2},\ldots,\alpha_{L}\}$ and \(0\leq \alpha_k \leq 1\). The backscattered signal from $k$-th AIoT device is
\begin{equation}
    \mathbf{x}_k[n]
    = \alpha_k r_k[n]
    =  \sqrt{P_t}\alpha_k  \mathbf{h}_{B,k}\mathbf{w}  s[n] .
\end{equation}

In practice, the modulation of AIoT devices could be controlled or uncontrolled. In this work, a controllable modulation is considered~\cite{7419631}, where the BS or receivers dynamically adjust the reflection pattern of AIoT devices. Meanwhile, the system could tolerate uncontrolled AIoT devices by letting each AIoT device randomly reflect according to a pattern known to the BS~\cite{7948789}.

\subsection{Receiver Model}

\subsubsection{Communication Receiver (User)}
Let \(\mathbf{h}_{B,U}\in\mathbb{C}^{1\times M_t}\) denote the BS--user channel. The received signal at the user at subcarrier \(n\) is
\begin{equation}
\begin{aligned}
     y_U[n]
    &= \underbrace{\mathbf{h}_{B,U}\mathbf{x}[n]}_{\text{Direct BS--user link}}
     + \sum_{k=1}^{K} \underbrace{h_{k,U} \mathbf{x}_k[n]}_{\text{ AIoT device relay}}
     + \mathbf z_U[n] \\
    &=  (
         \mathbf{h}_{B,U}
         + \sum_{k=1}^{K} \alpha_k h_{k,U}\mathbf{h}_{B,k}
        )\sqrt{P_t} \mathbf{w}  s[n]
       +  z_U[n],
\end{aligned}
\end{equation}
where \(z_U[n]\sim\mathcal{CN}(0,\sigma_U^2)\) is the additive white Gaussian noise (AWGN)~\cite{tse2005fundamentals}. Since \(\alpha_k\) is known to the legitimate system, the AIoT device-introduced signal term acts as a controllable relaying contribution to the equivalent communication channel. Instantaneous SINR received on the subcarrier \(n\) at the user is
\begin{equation}
    \gamma_U[n]
    = \frac{P_t |
          (\mathbf{h}_{B,U}
         + \sum_{k=1}^{K} \alpha_k h_{k,U}\mathbf{h}_{B,k} )\mathbf{w} |^2}{\sigma_U^2}.
    \label{eq:gamma_U_n}
\end{equation}

Then, the communication rate is calculated as
\begin{equation}
    R_{\text{c}}
    = \frac{1}{N}\sum_{n=0}^{N-1} \log_2 (1+\gamma_U[n] ).
    \label{eq:Rc}
\end{equation}

\subsubsection{Legitimate Sensing Receiver  (BS)}
The BS performs monostatic sensing of the target utilizing its $M_r$ receive antennas. Let $\mathbf{H}_{B,T}[n] = \mathbf{h}_{B , T}\mathbf{h}_{T , B}[n] \in \mathbb{C}^{M_r \times M_t}$ denote the round-trip channel matrix for the direct path (BS--target--BS), $\mathbf{G}_{B,k}[n] =\mathbf{h}_{B , k}\mathbf{h}_{k , B}[n] \in \mathbb{C}^{M_r \times M_t}$ denote the cascaded channel response for the direct backscatter link (BS--$k$-th AIoT device--BS), and $\mathbf{F}_{T,k}[n] =\mathbf{h}_{B , k}\mathbf{h}_{k , T} \mathbf{h}_{T , B}[n] \in \mathbb{C}^{M_r \times M_t}$ denote the cascaded channel for the AIoT device-assisted target reflection (BS--$k$-th AIoT device--target--BS), the received signal vector at the BS on subcarrier \(n\) is given by
\begin{equation}
\begin{aligned}
    \mathbf{y}_B[n]
    &= \underbrace{\mathbf{H}_{B,T}[n]\mathbf{x}[n]}_{\text{Direct target echo}}
     + \underbrace{\sum_{k=1}^{K} \alpha_k\mathbf{F}_{T,k}[n] \mathbf{x}[n]}_{\text{AIoT device-assisted target echo}} \\
    &+ \underbrace{\sum_{k=1}^{K} \alpha_k\mathbf{G}_{B,k}[n]  \mathbf{x}[n]}_{\text{AIoT device backscattered BS signal}}
     + \mathbf{z}_B[n],
\end{aligned}
\end{equation}
where \(\mathbf{z}_B[n]\sim\mathcal{CN}(\mathbf{0},\sigma_B^2\mathbf{I}_{M_r})\) denotes the AWGN at the BS. A practical channel model \cite{10570094} is applied where the BS infers the current sensing-channel CSI $\mathbf{H}_{B,T}[n]$ and 
$\mathbf{F}_{T,k}[n]$ from the previous-slot estimate and uses it for the current-slot design. However, the acquisition of perfect instantaneous CSI is practically infeasible due to estimation errors and the target mobility within the frame duration. Consequently, the imperfect CSI is modeled as follows \cite{10570094,10679658}:

\begin{equation}
\begin{aligned}
\mathbf{H}[n] &= \widehat{\mathbf{H}}[n] + \Delta\mathbf{H}[n],
\end{aligned}
\end{equation}
where $\widehat{\mathbf{H}}$ represents the estimated nominal channel of the channel $\mathbf{H}[n]$ derived from the previous time slot. $\Delta\mathbf{H}[n]$ denotes the channel state uncertainties that include estimation errors and time-varying mismatches, and
for an arbitrary matrix or vector channel $\mathbf{H}$, the normalized mean square error (NMSE) is calculated as
\begin{equation}
\text{NMSE} = \frac{\mathbb{E}\{\|\Delta\mathbf{H}\|^2\}}{\mathbb{E}\{\|\mathbf{H}\|^2\}} .
\end{equation}

Given the knowledge of \(\alpha_k\) and the estimated channel response \( \widehat{\mathbf{G}}_{B,k}[n]\)~\cite{8922800}, the BS reconstructs the ISAC signal backscattered by the AIoT devices and cancels it as follows 
\begin{equation}
     {\mathbf{y}}_{B,k}[n]
    = \sum_{k=1}^{K}  {\sqrt{P_t}\widehat {\mathbf{G}}}_{B,k}[n]\mathbf{w} \alpha_k s[n].
\end{equation}

Since the AIoT devices are controlled by the BS, the arrival times of the direct and backscattered signals can be made synchronous employing time offsets \cite{yang2017modulation}. Then, the received sensing signal after cancellation is
\begin{align}
    \widetilde{\mathbf{y}}_B[n]
    &= \mathbf{y}_B[n] -  {\mathbf{y}}_{B,k}[n] \nonumber\\
    &=\sqrt{P_t} \mathbf{H}_{B,T}[n]\mathbf{w}s[n] + \sqrt{P_t}\sum_{k=1}^K \mathbf{F}_{T,k}[n]\alpha_k\mathbf{w}s[n] \nonumber\\
    &  + \boldsymbol{\eta}_Bs[n]+{\mathbf{z}}_B[n],
\end{align}
where  $\boldsymbol{\eta}_B
  = \eta\sqrt{P_t}\sum_{k=1}^K  (\mathbf{G}_{B,k}[n]- \widehat{\mathbf{G}}_{B,k}[n] )\mathbf{w} \alpha_k$ denotes the residual signal assuming imperfect signal cancellation for the backscattered signal from AIoT devices at the BS, and $\eta$ denotes the imperfect AIoT device cancellation coefficient.

Subsequently, a linear combiner $\mathbf{v}_B[n]\in\mathbb{C}^{M_r\times 1}$ is applied to the received sensing signal for enhanced performance~\cite{tse2005fundamentals}. The resulting combined scalar output is
\begin{align}
  r_B[n]
  &= \mathbf v_B^H[n]\widetilde{\mathbf y}_B[n] \nonumber\\
  &= \mathbf v_B^H[n]\sqrt{P_t}(\mathbf H_{B,T}[n]
   + \sum_{k=1}^{K}\mathbf F_{T,k}[n]\alpha_k)\mathbf w s[n] \nonumber\\
  &+ \mathbf v_B^H[n]\boldsymbol{\eta}_B[n]s[n]
  + \mathbf v_B^H[n]{\mathbf z}_B[n]. 
  \label{BSSINR}
\end{align}

Based on the combined signal, the sensing SINR at the BS for subcarrier $n$ is derived as
\begin{equation}
\gamma_B[n] = \frac{{P_t}\left| \mathbf{v}_B^H[n] \left( \mathbf{H}_{B,T}[n] + \sum_{k=1}^K \alpha_k \mathbf{F}_{T,k}[n] \right) \mathbf{w} \right|^2}{\left| \mathbf{v}_B^H[n] \boldsymbol{\eta}_B[n] \right|^2 + \sigma_B^2 |\mathbf{v}_B[n]|^2},
\label{Sensing SINR}
\end{equation}
and the estimated sensing SINR at the BS can be denoted as $\widetilde{\gamma}_B[n]$, where the genuine channel fading $\mathbf{H}_{B,T}[n]$ and $\mathbf{F}_{T,k}[n]$ is replaced by $\widehat{\mathbf{H}}_{B,T}[n]$ and $\widehat{\mathbf{F}}_{T,k}[n]$.




\subsection{Sensing Eavesdropper Model (SEve)}
The sensing eavesdropper, SEve, is a passive bistatic sensing receiver with no prior knowledge of the transmit signals and requires a reference signal to demodulate the surveillance signals from targets.
Let \(\mathbf{H}_{B,E}[n] \in\mathbb{C}^{M_e\times M_t}\) denote the BS--SEve channel. We consider that SEve is equipped with a large number of receiving antennas or separately deployed directional antennas, and is aware of the location of the BS, thus it can perfectly separate the reference signal from the BS and other signals \cite{zhang2024covert}. Then, the reference signal received directly from the BS to SEve is 
\begin{equation}
    \mathbf{y}_{E,\text{ref}}[n]
    = \sqrt{P_t}\mathbf{H}_{B,E}[n]\mathbf{w}  s[n] + \mathbf{z}_{E,\text{ref}}[n],
\end{equation}
where $\mathbf{z}_{E,\text{ref}}[n]\sim\mathcal{CN}(\mathbf{0},\sigma_{E,\text{ref}}^2\mathbf{I}_{M_e})$ is the AWGN.

Let $\mathbf u_{E,\mathrm{ref}}[n]\in\mathbb C^{M_e\times 1}$ be the linear combiner applied to
$\mathbf y_{E,\mathrm{ref}}[n]=\mathbf H_{B,E}[n]\mathbf w s[n]+\mathbf z_{E,\mathrm{ref}}[n]$. The  scalar reference after the combiner is
\begin{align}
  r_{E,\mathrm{ref}}[n]
  &= \mathbf u_{E,\mathrm{ref}}^{H}[n]\mathbf y_{E,\mathrm{ref}}[n] \nonumber \\
  & = \mathbf u_{E,\mathrm{ref}}^{H}[n]\sqrt{P_t}\mathbf H_{B,E}[n]s[n]\mathbf w
     + \nu_{E,\mathrm{ref}}[n],
     \label{15}
\end{align}
where $\nu_{E,\mathrm{ref}}[n]= \mathbf u_{E,\mathrm{ref}}^{H}[n]\mathbf z_{E,\mathrm{ref}}[n]$.

The SEve could estimate the channel between the BS and SEve, i.e., 
${c}_E[n] = \mathbf u_{E,\mathrm{ref}}^{H}[n]\sqrt{P_t}\mathbf H_{B,E}[n]\mathbf w$ using pilot signals. Then, SEve can obtain a noisy estimate of the transmitted symbol as
\begin{equation}
    \tilde{s}_E[n]
    = \frac{r_{E,\mathrm{ref}}[n]}{{c}_E[n]}
    = s[n] + \tilde{z}_{E,\mathrm{ref}}[n],
\end{equation}
where $\tilde{z}_{E,\mathrm{ref}}[n]$ shars the same statistical properties as ${z}_{E,\mathrm{ref}}[n]$.

Let \(\mathbf{H}_{T,E}[n] = \mathbf{h}_{B,T}\mathbf{h}_{T,E}[n] \in\mathbb{C}^{M_e\times M_t}\) denote the BS--target--SEve channel, \(\mathbf{G}_{k,E}[n] = \mathbf{h}_{B,k}\mathbf{h}_{k,E}\in\mathbb{C}^{M_e\times M_t}\) the BS--$k$-th AIoT device--SEve channel, and \(\mathbf{Q}_{k,E}[n] = \mathbf{h}_{B,k}\mathbf{h}_{k,T}\mathbf{h}_{T,E}\in\mathbb{C}^{M_e\times M_t}\) the BS--$k$-th AIoT device--target--SEve sensing channel. The eavesdropped sensing echo signal at SEve is
\begin{equation}
\begin{aligned}
    \mathbf{y}_E[n]
    &=\underbrace{ \sqrt{P_t}\mathbf{H}_{T,E}[n]\mathbf{w}  s[n]}_{\text{Target echo}}
     + \underbrace{\sqrt{P_t}\sum_{k=1}^{K} \mathbf{G}_{k,E}[n]\mathbf{w} \alpha_k s[n]}_{\text{AIoT device backscattered Signal}} \\
     & + \underbrace{\sqrt{P_t}\sum_{k=1}^{K} \mathbf{Q}_{k,E}[n]\mathbf{w} \alpha_k s[n]}_{\text{AIoT device-assisted target echo}}
     + \mathbf{z}_{E,T}[n],
\end{aligned}
\end{equation}
where \(\mathbf{z}_{E,T}[n]\sim\mathcal{CN}(\mathbf{0},\sigma_{E,T}^2\mathbf{I}_{M_e})\). Since SEve does not know about the backscatter modulation of AIoT devices \cite{zhang2025artificial}, the backscattered signal from the BS to SEve serves as interference.

With a linear combiner \(\mathbf u_E[n]\in\mathbb C^{M_e\times 1}\), the combined scalar sensing observation at SEve on subcarrier \(n\) is
\begin{align}
r_E[n]
&= \mathbf u_E^{H}[n]\mathbf y_E[n] \nonumber\\
&= \mathbf u_E^{H}[n]\sqrt{P_t}(\mathbf H_{T,E}[n]
 +\sum_{k=1}^{K} \alpha_k\mathbf G_{k,E}[n]  \nonumber
 \\ &+ \sum_{k=1}^{K}\alpha_k\mathbf Q_{k,E}[n])\mathbf w   s[n]
 + \mathbf u_E^{H}[n]\mathbf z_{E,T}[n].
\label{eq:rE_combined}
\end{align}

Using the eavesdropped reference $\tilde s_E[n]$, SEve typically uses a matched filter (MF) to obtain the range profile of the target signal~\cite{10587082}
\begin{align}
  \tilde r_E[n]
  &= r_E[n]\;\tilde s_E^{*}[n].
\end{align}

Then, the SINR for the detection of SEve is derived in \eqref{eq:sinr_eve_sensing}, where \(\tilde \sigma_{E,\mathrm{ref}}^{2}\) captures the effective residual distortion induced by SEve's estimation error in \(\tilde s_E[n]\) and is interpreted as an additional equivalent noise term at the MF output.

\begin{figure*}[!t]
\begin{equation}
\gamma_{E}[n]
=
\frac{{P_t}
\left|
\mathbf u_E^{H}[n]\mathbf H_{T,E}[n]\mathbf w
\right|^{2}
}{P_t
\left|
\mathbf u_E^{H}[n]\sum_{k=1}^{K}\mathbf G_{k,E}[n]\mathbf w \alpha_k
\right|^{2}
+P_t
\left|
\mathbf u_E^{H}[n]\sum_{k=1}^{K}\mathbf Q_{k,E}[n]\mathbf w \alpha_k
\right|^{2}
+\sigma_{E,T}^{2}\left\|\mathbf u_E[n]\right\|^{2}
+\tilde \sigma_{E,\mathrm{ref}}^{2}
}.
\label{eq:sinr_eve_sensing}
\end{equation}
\hrulefill
\end{figure*}

\begin{figure}[tp]
    \centering
    \includegraphics[width=\linewidth]{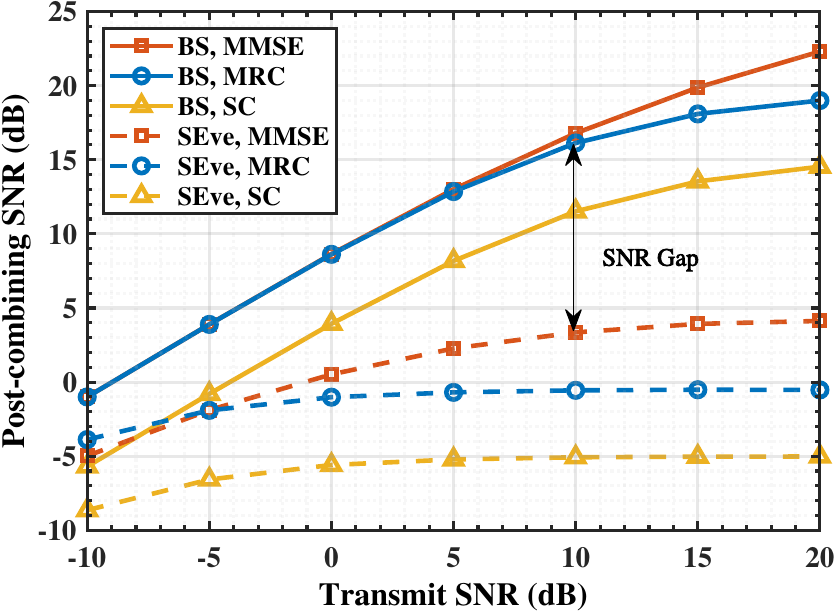}
    \caption{Sensing SINR performance versus transmit SNR for different linear combiners at the BS and the passive eavesdropper under 10 AIoT devices}
    \label{2}
\end{figure}

\textbf{Remark 1: Combiner Selection and Security Analysis.}  
Figure \ref{2} compares the post-combining sensing SINR performance of the BS and SEve under 10 AIoT devices, using three standard linear combining schemes~\cite{9737357,sengijpta1995fundamentals}, including minimum mean square error (MMSE), maximum ratio combining (MRC), and selection combining (SC). Two critical observations can be derived from these results. 
First, the MMSE combiner consistently outperforms both MRC and SC schemes for both the BS and SEve, particularly in the high-SNR regime. This superiority comes from the fact that MMSE is the optimal strategy to maximize SINR detection, as it effectively balances noise suppression with interference mitigation induced by AIoT devices~\cite{sengijpta1995fundamentals}. Consequently, MMSE is adopted as the combining strategy for both parties in the subsequent analysis. Second, the figure reveals a significant SNR gap of approximately 13 dB between the BS and Sve at a transmit SNR of 10 dB. A comparison of the SINR expressions in \eqref{BSSINR} and \eqref{eq:sinr_eve_sensing} indicates that this disparity is the result of an asymmetry in system knowledge. Specifically, BS knows the reflection coefficients of AIoT devices, $\alpha_k$, and can estimate the relevant CSI~\cite{8746230}, to reconstruct and cancel the backscattered interference. In contrast, Sve lacks this critical information. Thus, the signals reflected by the AIoT devices manifest as strong, unmitigated active interference, severely degrading her sensing performance.

\section{Problem Formulation}\label{sec:SOP}

The core strategy of AmbSentry is to leverage AIoT devices to generate artificial clutter that obscures SEve's detection and estimation of the true target.
This is accomplished by shaping the zero-Doppler ambiguity function of the reflected signal from the target and the AIoT devices to produce significant ambiguous peaks in the range profile for SEve, degrading its detection and estimation performance. The subsequent section defines the ambiguity function, derives the closed-form ISL as a metric to quantify the security performance against SEve, and formulates an optimization problem to maximize sensing security.

\subsection{Ambiguity Function}
In practice, SEve’s instantaneous channels and receiver processing are generally unavailable, making SEve’s CSI-dependent formulation intractable for the design of transmit ISAC waveforms and AIoT device signaling. Therefore, our SEve-agnostic approach quantifies and controls the shape of the MF range profile through an ambiguity function that retains only the controllable components contributed by the target-reflected echo and the AIoT device-backscattered signals superimposed on the known ISAC waveform. The frequency-domain auto-correlation function (ACF) of the combined signal is calculated as~\cite{jankiraman2018fmcw}
\begin{equation}
 {\Lambda}
= \sqrt{N}\,\mathbf F_{N}^{H} (\mathbf S^{2}\mathbf Q )\mathbf 1_{N},
\end{equation}
where $\mathbf F_{N}^{H}$ denotes the normalized DFT matrix of size $N$, and $\mathbf S^{2}=\mathrm{diag}(|s[0]|^{2},\dots,|s[N-1]|^{2})$ collects the per-subcarrier symbol energies of the ISAC waveform, and
$\mathbf Q=\mathrm{diag}(q[0],\dots,q[N - 1])$ is the combined signal  matrix transmitted to SEve. Under this SEve CSI-independent characterization, $q[n]$ is defined as the composite per-subcarrier coefficient of the target-reflection and AIoT device-backscatter superposition that governs the resulting delay-domain profile, and it is used as the basis for the ACF computation as follows
\begin{equation}
q[n]=
\begin{cases}
    \sqrt{P_t}\mathbf  \alpha_t  \widehat{\mathbf{h}}_{B,T}[n]\mathbf w, & K=0,\\[6pt]
 \sqrt{P_t} \mathbf  \alpha_t \widehat{\mathbf{h}}_{B,T}[n]\mathbf w  + \sqrt{P_t} \sum_{k=1}^{K}\alpha_k \mathbf h_{B,k}[n]\mathbf w, 
    & K\neq 0,
\end{cases}
\label{eq:q_cases}
\end{equation}
where   $\alpha_t \sim \mathcal{CN}(0, \sigma_t^2)$ \cite{bicua2019multicarrier} represents the radar cross section (RCS) of the sensing target, $\widehat{\mathbf{h}}_{B,T}[n]$ denotes the imperfect channel estimation of the channel from the BS to the target, and the $\nu$-th element is expanded in scalar form as
\begin{equation}
  \Lambda[\nu]
  = \sum_{n=0}^{N-1} |s[n]|^{2}  q[n]   e^{j\frac{2\pi}{N}\nu n}.
\end{equation}

Let $\widetilde{\Lambda}[\nu]$ denote an ideal secure ACF that creates nonuniform artificial targets at AIoT device-induced discrete delays $\ell_k$. Then, the desired ACF for secure sensing in the power domain can be represented as
\begin{equation}
  \widetilde{\Lambda}[\nu]
   = 
 \underbrace{ P_t||\widehat{\mathbf{h}}_{B,T} \mathbf w \|^{2} \delta[\nu]}_{\text{Mainlobe}}
   + \underbrace{ P_t\sum_{k=1}^{K}|\alpha_k|^{2}   \|{\mathbf h}_{B,k}\mathbf w \|^{2} \delta[\nu-\ell_{k}]}_{\text{Artificial peaks}},
  \label{eq:secure_acf_nonperiodic}
\end{equation}
with OFDM bandwidth $B$ and spacing $\Delta f=B/N$, the maximum unambiguous range is $R_{\max}=\frac{cN}{2B}$. 
A discrete lag $\ell_k$ maps to range $R_k=\frac{c}{2B} \ell_k$ with round-trip delay $\tau_k=\frac{\ell_k}{N\Delta f}$. 

\textbf{Remark 2: Ideal Secure ACF.}  Equation \eqref{eq:secure_acf_nonperiodic} characterizes the ideal sensing ambiguity function observed by the eavesdropper. The response comprises two distinct components: the mainlobe corresponding to the true target's echo, and the artificial sidelobe induced by the AIoT devices. By jointly optimizing the transmit beamforming coefficients $\mathbf{w}$ and the reflection coefficients of the AIoT device, $\alpha_k$, the system aims to maximize the power of these artificial peaks relative to SEve's mainlobe. Consequently, SEve perceives multiple high-intensity false targets, rendering it unable to distinguish the true target range from the AIoT devices, thereby effectively safeguarding sensing security~\cite{11202391}.

\subsection{Security Performance: Integrated Sidelobe Level}
\label{subsec:isl_diff}
The efficacy of AIoT device jamming against the eavesdropper is quantified using an ISL metric that normalizes the sidelobe energy introduced by the AIoT device to the mainlobe energy of the signal transmitted from the source and reflected by the target. The ISL is calculated as
\begin{equation}
\Delta\mathrm{ISL} 
=
\frac{\sum_{k \in \mathcal{K}} \mathbb{E}\{|\Lambda[\ell_k]|^2\}}{\mathbb{E}\{|\Lambda[0]|^2\}}.
\label{eq:ISL_diff_def_en}
\end{equation}

Since the residual cancellation errors of different AIoT devices originate from different scattering paths, their phase and delay errors are typically independent or distinguishable. Therefore, the cross-terms tend toward zero in the statistical average sense. Consequently, this paper adopts a non-coherent residual-power model. For each delay component $\mathbb{E} \{ |\Lambda[\nu] |^{2} \}$, it is calculated as
\begin{align}
\mathbb{E} \{ |\Lambda[\nu] |^{2} \} \nonumber &=  
\sum_{n=0}^{N-1}\sum_{m=0}^{N-1}
\mathbb{E} \{|s_n|^{2}|s_m|^{2} \} 
q[n] q[m]^{*} e^{j\frac{2\pi}{N}\nu(n-m)} \nonumber\\
&=  (
\mu_2^{2}  |\sum_{n=0}^{N-1} q[n] e^{j\frac{2\pi}{N}\nu n}  |^{2}
+ (\mu_{4}-\mu_{2}^{2} )\sum_{n=0}^{N-1} |q[n] |^{2}
 ).
\label{eq:key_second_moment}
\end{align}


To derive the metric for generating specific interference energy, we first characterize the mainlobe power, which is given by
\begin{align}
\mathbb{E}\{|\Lambda [0]|^2\} &= \mu_2^{2} \left| \sum_{n=0}^{N-1} q[n] \right|^2+ (\mu_4 - \mu_2^2) \sum_{n=0}^{N-1} |q[n]|^2.
\label{eq:mainlobe_power}
\end{align}

Unlike conventional ISL that integrates sidelobes over the entire delay domain, the proposed deception strategy focuses on maximizing the energy at specific delays corresponding to the AIoT devices. Let $\mathcal{L} = \{\ell_1, \ell_2, \dots, \ell_K\}$ denote the set of discrete delay indices associated with the $K$ AIoT devices. The sum expected power of the ambiguity function, specifically over these delays $\ell_k \in \mathcal{L}$ in the sidelobe, is calculated as
\begin{align}
& \sum_{k=1}^{K} \mathbb{E} \left\{ |\Lambda[\ell_k]|^{2} \right\} \nonumber \\
&= \sum_{k=1}^{K} \left( \mu_2^{2} \left| \sum_{n=0}^{N-1} q[n] e^{j\frac{2\pi}{N}\ell_k n} \right|^{2} + (\mu_{4}-\mu_{2}^{2} ) \sum_{n=0}^{N-1} |q[n] |^{2} \right). \nonumber \\
&= \mu_2^{2} \sum_{k=1}^{K} \left| \sum_{n=0}^{N-1} q[n] e^{j\frac{2\pi}{N}\ell_k n} \right|^{2}
+ K (\mu_{4}-\mu_{2}^{2} ) \sum_{n=0}^{N-1} |q[n] |^{2}
\label{eq:sidelobe_power}
\end{align}

Finally, by substituting \eqref{eq:sidelobe_power} and \eqref{eq:mainlobe_power} into the ISL definition \eqref{eq:ISL_diff_def_en}, the closed-form ISL expression is derived as
\begin{align}
&\Delta \text{ISL}= \nonumber \\
&\frac{\mu_2^2 \sum_{k=1}^{K} \left| \sum_{n=0}^{N-1} q[n] e^{j\frac{2\pi}{N}\ell_k n} \right|^2 + K (\mu_4 - \mu_2^2) \sum_{n=0}^{N-1} |q[n]|^2}{\mu_2^2 \left| \sum_{n=0}^{N-1} q[n] \right|^2 + (\mu_4 - \mu_2^2) \sum_{n=0}^{N-1} |q[n]|^2}.
\label{ISL}
\end{align}

\subsection{Secure AIoT device-assisted ISAC System Design}\label{subsec:secure_design}

This paper focuses on secure sensing in AmbSentry, where BS cooperatively designs the transmit beamforming and the modulation of AIoT devices to optimize sensing security. Specifically, the corresponding optimization problem is to maximize the ISL under S\&C performance constraints, which is formulated as follows
\begin{subequations}\label{eq:main_problem}
\begin{align}
\text{(P1)} \max_{\mathbf w,\  {\alpha}}\quad 
& \Delta\mathrm{ISL} 
\label{eq:obj_isl} \\[1mm]
\text{s.t.}\quad
& \frac{1}{N}\sum_{n=0}^{N-1}\log_2  (1+\gamma_U[n] )\ \ge\ R_{\min},
\label{eq:comm_qos} \\[1mm]
& \widetilde{\gamma}_B(\mathbf w, {\alpha})\ \ge\ \Gamma_{\min},
\label{eq:sens_qos} \\[1mm]
& \alpha_k\in\mathcal A_k,\quad k=1,\ldots,K,
\label{eq:bd_discrete}
\end{align}
\end{subequations}
where problem (P1) maximizes the ISL by jointly shaping the transmitted signal via the BS beamformer $\mathbf{w}$ and the AIoT device reflection coefficients ${\alpha}$. The constraint \eqref{eq:comm_qos} guarantees a minimum average spectral efficiency $R_{\min}$ for the legitimate user, ensuring reliable communication. Similarly, the quality of the detection in the BS is preserved by constraint \eqref{eq:sens_qos}, which enforces a lower bound $\Gamma_{\min}$ in the post-processing detection SINR. Furthermore, constraint \eqref{eq:bd_discrete} restricts the reflection coefficients to a discrete set $\mathcal{A}_k$ to account for practical hardware limitations.

Given the coupled nature of the beamforming vector and modulation decisions of AIoT devices, an inherent trade-off exists between maximizing sensing security and maintaining robust S\&C performance. Characterizing this trade-off is critical for guiding the practical deployment of AIoT device-assisted secure ISAC systems. However, the original problem (P1) is non-convex due to three primary factors. First, the objective function aims to maximize the ISL, which is formulated as a fractional programming problem (i.e., maximizing a ratio $\Delta\mathrm{ISL} = \frac{\mathcal{N}(\mathbf{w}, \boldsymbol{\alpha})}{\mathcal{D}(\mathbf{w}, \boldsymbol{\alpha})}$). Such fractional objectives are inherently non-convex because the Hessian matrix of a ratio of functions is generally not negative semi-definite for a maximization problem, making global optimality difficult to achieve. Second, there is a strong coupling between the beamformer $\mathbf{w}$ and the AIoT device modulations $\boldsymbol{\alpha}$ in the ISL expression. Third, the feasible set for the reflection coefficients $\boldsymbol{\alpha}$ is discrete, restricting the values to a finite set of amplitudes or phases. This imposes combinatorial constraints, rendering the problem an integer programming problem.
Therefore, (P1) is highly NP-hard~\cite{gao2023cooperative}. To solve this problem, an efficient optimization algorithm is presented in the following.

\section{The Proposed Optimization Algorithm for Sensing Security}
\label{sec:opt_pathA}

To address the non-convex Problem in (P1), we develop a joint design of the BS transmit beamformer $\mathbf w$ and the $K$ AIoT devices' reflection strategy $ {\boldsymbol \alpha}=\{\alpha_k\}_{k=1}^{K}$ to maximize the sensing–security metric $\Delta\mathrm{ISL}$. Specifically, a generalized Dinkelbach transformation is first adopted to address the fractional structure of the objective function. Then, a BCD procedure is applied to alternately optimize the beamforming vector and the AIoT device backscatter using the SCA and SDR methods for non-convex relaxation.

\subsection{Generalized Dinkelbach Transformation}
\label{sec:mm_transform}

Define an auxiliary variable $\lambda$, in the $t$-th outer iteration, (P1) is transformed to the following subproblem~\cite{dinkelbach1967nonlinear}:
\begin{align}
\text{(P2)} \quad & \arg \max_{\mathbf{w}, \boldsymbol{\alpha}} \quad \mathcal{F}(\mathbf{w}, \boldsymbol{\alpha} | \lambda^{(t)}) = \mathcal{N}(\mathbf{w}, \boldsymbol{\alpha}) - \lambda^{(t)} \mathcal{D}(\mathbf{w}, \boldsymbol{\alpha}), \nonumber \\
    \text{s.t.} & \quad \eqref{eq:comm_qos} - \eqref{eq:bd_discrete},
\label{eq:dinkelbach_obj}
\end{align}
where  $\lambda^{(t)}$ is the auxiliary variable updated at the $t$-th iteration and $\mathcal{F}(\mathbf{w}, \boldsymbol{\alpha} | \lambda^{(t)}) \ge 0$ is ensured by the proposed BCD procedure outlined in the next subsection \ref{BCD}.

Upon obtaining the solution $(\mathbf{w}^{(t)}, \boldsymbol{\alpha}^{(t)})$ from the inner loop, the auxiliary variable is updated as 
\begin{equation}
\lambda^{(t+1)} = \frac{\mathcal{N}(\mathbf{w}^{(t)}, \boldsymbol{\alpha}^{(t)})}{\mathcal{D}(\mathbf{w}^{(t)}, \boldsymbol{\alpha}^{(t)})},
\label{30}
\end{equation}
which continues until the convergence criterion $|\mathcal{N}(\mathbf{w}, \boldsymbol{\alpha}) - \lambda \mathcal{D}(\mathbf{w}, \boldsymbol{\alpha})| \le \epsilon$ is met.

\subsection{Problem Decomposition}  \label{BCD}
The problem (P2) is still non-convex due to coupling between the beamforming vector and discrete AIoT device modulations, making it NP-hard to solve optimally. To overcome this, a BCD framework is adopted to decompose (P2) into two subproblems.

    \subsubsection{ Beamformer Optimization} With fixed reflection coefficients $ {\alpha}$, problem (P2) is reduced to the following formulation 
 \begin{align}
    \text{(P3)} \quad \max_{\mathbf{w}} \quad & \mathcal{F}(\mathbf{w}) \nonumber \\
    \text{s.t.} & \quad \eqref{eq:comm_qos}, \eqref{eq:sens_qos}.
    \label{eq:subproblem_W}
    \end{align}

    \subsubsection{Reflection Coefficient Optimization} With fixed beamformer $\mathbf{w}$, the optimization of $ {\boldsymbol \alpha}$ is formulated as 
    \begin{align}
    \text{(P4)} \quad \max_{ {\alpha}} \quad & \mathcal{F}( {\boldsymbol \alpha}) \nonumber \\
    \text{s.t.} & \quad \eqref{eq:comm_qos} - \eqref{eq:bd_discrete}.
    \label{eq:subproblem_alpha}
    \end{align}

    The proposed algorithm iteratively updates $\mathbf{w}$ and $ { {\alpha}}$ by solving (P3) and (P4) iteratively. Under the continuous relaxation and assuming each convexified subproblem is solved exactly and yields a non-decreasing surrogate objective, the proposed BCD-SCA procedure converges to a stationary point of the relaxed surrogate problem \cite{dinkelbach1967nonlinear}.

\subsection{Block Update Strategy}
\subsubsection{Block Update on $\mathbf{w}$}
Fixing ${\alpha}$, an SDR method is employed to optimize $\mathbf{w}$. Define $\mathbf{W} = \mathbf{w}\mathbf{w}^H \succeq \mathbf{0}$.
The objective function in \eqref{eq:subproblem_W} is linear with respect to $\mathbf{W}$. Specifically, neglecting constant terms, the optimized objective in (P3) can be rewritten as
\begin{equation}
\mathcal{F}_{\mathbf{w}}(\mathbf{W}) = \mathrm{tr} ( \mathbf{\Phi}^{(t)} \mathbf{W}),
\label{eq:obj_W}
\end{equation}
where the aggregation matrix $\mathbf{\Phi}^{(t)}$ is expressed as
\begin{equation}
\mathbf{\Phi}^{(t)}=\sum_{n=0}^{N-1} g^{(t)}[n]\,
\widetilde{\mathbf H}_{B}^{H}[n]\widetilde{\mathbf H}_{B}[n],
\label{34}
\end{equation}
where $\widetilde{\mathbf H}_{B}[n] = \mathbf{h}_{B,T}[n] + \sum_{k=1}^{K}\alpha_k \mathbf{h}_{B,k}[n]$ denoting the composite coefficient matrix and $g^{(t)}[n]$ denoting the effective subcarrier-dependent weight at the $t$-th Dinkelbach iteration, induced by the closed-form expression of ISL in \eqref{ISL} and the auxiliary parameter $\lambda^{(t)}$.

From \eqref{eq:gamma_U_n}, the instantaneous receive SINR on subcarrier $n$ at the user can be written as
\begin{equation}
\gamma_U[n] = \frac{ P_t|\mathbf{h}_U[n]\mathbf{w} |^2}{\sigma_U^2}
           = P_t\sigma_U^{-2}\,\mathrm{tr}  (\mathbf{h}_U^{H}[n]\mathbf{h}_U[n]\mathbf{W} ),
\end{equation}
where $\mathbf{h}_U[n] = \mathbf{h}_{B,U}[n] + \sum_{k=1}^{K}\alpha_k\,h_{k,U}[n]\mathbf{h}_{B,k}[n]$ is the effective communication channel on subcarrier $n$.
Accordingly, the communication rate constraint in \eqref{eq:comm_qos} is reformulated as
\begin{equation}
\sum_{n=0}^{N-1} \log_2   \left( 1 + P_t\sigma_U^{-2}\,\mathrm{tr}  (\mathbf{h}_U^{H}[n]\mathbf{h}_U[n]\mathbf{W} ) \right) \ge N R_{\min},
\label{eq:rate_convex_W}
\end{equation}
which is the superlevel set of a concave function.

To handle the sensing SINR constraint in \eqref{eq:sens_qos}, we first define
the effective sensing channel and residual interference channel as $\widetilde{\mathbf G}_{B}[n]= \sum_{k=1}^{K} \alpha_k
\left(\mathbf G_{B,k}[n]-\widehat{\mathbf G}_{B,k}[n]\right)$. Then, the auxiliary matrices are given by
\begin{subequations}
\begin{align}
\mathbf R_B[n]
&= \widetilde{\mathbf H}_{B}^{H}[n]\mathbf v_B[n]\mathbf v_B^{H}[n]
   \widetilde{\mathbf H}_{B}[n], \\
\mathbf R_{\eta,B}[n]
&= \widetilde{\mathbf G}_{B}^{H}[n]\mathbf v_B[n]\mathbf v_B^{H}[n]
   \widetilde{\mathbf G}_{B}[n].
\end{align}
\end{subequations}
Accordingly, the desired sensing power and the residual
AIoT-induced interference power can be expressed as
\begin{subequations}
\begin{align}
\left|\mathbf v_B^{H}[n]\widetilde{\mathbf H}_{B}[n]\mathbf w\right|^2
&= \operatorname{tr}\left(\mathbf R_B[n]\mathbf W\right), \\
\left|\mathbf v_B^{H}[n]\boldsymbol{\eta}_B[n]\right|^2
&= \operatorname{tr}\left(\mathbf R_{\eta,B}[n]\mathbf W\right).
\end{align}
\end{subequations}

Thus, the minimum sensing SINR constraint can be reformulated as
\begin{align}
&P_t\sum_{n=0}^{N-1} \mathrm{tr}(\mathbf{R}_B[n] \mathbf{W})
~\ge~ \nonumber \\
&\Gamma_{\min} \left(P_t
\sum_{n=0}^{N-1}\mathrm{tr}(\mathbf{R}_{\eta,B}[n]\mathbf{W})
+\sum_{n=0}^{N-1}\sigma_B^{2}\|\mathbf{v}_B[n]\|_2^2
\right),
\label{eq:sens_W}
\end{align}
which is convex in $\mathbf{W}$.

Consequently, the $\mathbf{w}$-subproblem is cast as a convex SDP:
\begin{equation} \label{P5}
\text{(P5)}~~
\max_{\mathbf{W} \succeq \mathbf{0}}~~ \mathcal{F}_{\mathbf{w}}(\mathbf{W})
\quad \text{s.t.} \quad
\eqref{eq:rate_convex_W},~ \eqref{eq:sens_W},~ \mathrm{tr}(\mathbf{W}) \le 1,
\end{equation}
which can be efficiently solved using the CVX toolbox~\cite{cvx}.
Note that if the optimal solution $\mathbf{W}^\star$ is rank-one, $\mathbf{w}^\star$ is directly obtained from its principal eigenvector.
Otherwise, a standard Gaussian randomization \cite{1634819} is applied to extract a feasible beamformer $\mathbf{w}^\star$.

\subsubsection{Block Update on $ {\alpha}$}

With the transmit beamformer $\mathbf{w}$ held constant, the backscatter modulation  $\boldsymbol{\alpha}$ are optimized. To address the intractability arising from the discrete constraint $\alpha_k \in \mathcal{A}_k$, a continuous relaxation is adopted, where $0 \le \alpha_k \le 1$.

In the $t$-th iteration, define the effective vectors $\mathbf{z}_0[n]= \mathbf{H}_{B,T}[n]\mathbf{w}$ and $\mathbf{z}_k[n]= \mathbf{H}_{B,k}[n]\mathbf{w}$. Then, the optimized objective becomes
\begin{equation}
\mathcal{F}_{{\alpha}}({\alpha})
=
\sum_{n=0}^{N-1} g^{(t)}[n]\,
 \|
\mathbf{z}_0[n]+\sum_{k=1}^{K}\alpha_k \mathbf{z}_k[n]
 \|^2 .
\label{eq:obj_alpha_sum}
\end{equation}

Define $\mathbf{Z}[n]=[\mathbf{z}_1[n],\ldots,\mathbf{z}_K[n]]$, \eqref{eq:obj_alpha_sum} is rewritten as a convex quadratic function:
\begin{equation}
\mathcal{F}_{{\alpha}}({\alpha})
~=~
{\alpha}^{T}\mathbf{C}^{(t)}{\alpha}
+2(\mathbf{d}^{(t)})^{T}{\alpha},
\label{eq:obj_alpha_quad}
\end{equation}
where $\mathbf{C}^{(t)} = \sum_{n=0}^{N-1} g^{(t)}[n] \Re \{\mathbf{Z}^{H}[n]\mathbf{Z}[n]\}$ and $\mathbf{d}^{(t)} = \sum_{n=0}^{N-1} g^{(t)}[n] \Re \{\mathbf{Z}^{H}[n]\mathbf{z}_0[n]\}$.

Since maximizing a convex function is non-convex, the SCA method is applied by using 
the first-order Taylor expansion at $\boldsymbol{\alpha}^{(t)}$ as a global linear lower bound
\begin{equation}
\widetilde{\mathcal{F}}_{\alpha}(\boldsymbol{\alpha}) = 2 (\boldsymbol{\alpha}^{(t)})^T \mathbf{C}^{(t)} \boldsymbol{\alpha} - (\boldsymbol{\alpha}^{(t)})^T \mathbf{C}^{(t)} \boldsymbol{\alpha}^{(t)} + 2 (\mathbf{d}^{(t)})^T \boldsymbol{\alpha}.
\label{eq:alpha_obj_sca}
\end{equation}

For the communication rate constraint, the effective scalar channel is $x_n( {\alpha}) = \mathbf{h}_{B,U}[n]\mathbf{w} + \sum_{k=1}^{K} \alpha_kh_{k,U}[n]\mathbf{h}_{B,k}[n]\mathbf{w}$. The convexity of $|x_n(\boldsymbol{\alpha})|^2$ allows for a linear lower bound
\begin{equation}
|x_n( {\alpha}) |^2 \ge \mathcal{L}_n(\boldsymbol{\alpha}) = 2\Re \{x_n( {\alpha}^{(t)})^{ *}\,x_n( {\alpha}) \} - |x_n( {\alpha}^{(t)}) |^2.
\label{eq:xn_mag_sq_lb}
\end{equation}

Consequently, the non-convex rate constraint is replaced by a convex subset
\begin{equation}
\sum_{n=0}^{N-1} \log_2 \left( 1 +P_t \sigma_U^{-2} \mathcal{L}_n(\boldsymbol{\alpha}) \right) \ge N R_{\min}.
\label{eq:rate_convex_alpha}
\end{equation}


For the sensing constraint, we explicitly preserve the dependence of the residual AIoT-induced interference on $\boldsymbol{\alpha}$. Define $y_n(\boldsymbol{\alpha}) = \mathbf{v}_B^H[n]\mathbf{H}_{B,T}[n]\mathbf{w}
\sum_{k=1}^{K}\alpha_k \mathbf{v}_B^H[n]\mathbf{F}_{T,k}[n]\mathbf{w}$, and $ b_n(\boldsymbol{\alpha}) =
  \mathbf{v}_B^H[n]\sum_{k=1}^{K}\alpha_k
  \left(\mathbf{G}_{B,k}[n]-\hat{\mathbf{G}}_{B,k}[n]\right)\mathbf{w}$.
 Then, the sensing SINR constraint can be written as
  $|y_n(\boldsymbol{\alpha})|^2 \ge \Gamma_{\min}\left(|b_n(\boldsymbol{\alpha})|^2+\sigma_B^2|\mathbf{v}_B[n]|^2\right)$. By applying SCA to the left-hand side, a convex inner approximation is obtained as
\begin{align}
&2\Re{y_n(\boldsymbol{\alpha}^{(t)})^{*}y_n(\boldsymbol{\alpha}}
  -|y_n(\boldsymbol{\alpha}^{(t)})|^2
  \ge \\
  &\Gamma_{\min}\left(|b_n(\boldsymbol{\alpha})|^2+\sigma_B^2|\mathbf{v}_B[n]|^2\right).
  \label{eq:sensing_sca_linear}
\end{align}

Combining \eqref{eq:alpha_obj_sca}, \eqref{eq:rate_convex_alpha}, and \eqref{eq:sensing_sca_linear}, the SCA-convexified $ {\alpha}$-subproblem is formulated as
\begin{subequations}
\label{eq:alpha_LP}
\begin{align}
\text{(P6)}\quad
\max_{ {\alpha}}~~
& \widetilde{\mathcal{F}}_{\alpha}(\boldsymbol{\alpha})
\label{eq:alpha_LP_obj}
\\
\text{s.t.}~~
& 0 \le \alpha_k \le 1,\quad \forall k,
\label{eq:alpha_LP_box}
\\
& \eqref{eq:rate_convex_alpha}, \eqref{eq:sensing_sca_linear}.
\label{eq:alpha_LP_con}
\end{align}
\end{subequations}
Problem \eqref{eq:alpha_LP} is a convex optimization problem and can be efficiently solved by the CVX~\cite{cvx}.


\begin{algorithm}[!t]
\footnotesize
\SetCommentSty{small}
\LinesNumbered
\caption{Joint Beamforming and AIoT device Modulation Design Algorithm}
\label{alg:ISL_Maximization}
\KwIn{Channel information $\mathbf{H}_{q , p}$, the  covariance matrices $\mathbf{R}_k$, $P_t$, $R_{\min}$, $\Gamma_{\min}$, the discrete set $\mathcal{A}_k$, and convergence thresholds $\epsilon_{out}, \epsilon_{1}, \epsilon_{2} = 10^{-3}$} 

\KwOut{Optimal beamformer $\mathbf{w}$ and reflection coefficients $\boldsymbol{\alpha}$.}

Initialize beamformer $\mathbf{w}^{(0)}$, reflection coefficients $\boldsymbol{\alpha}^{(0)}$.\\
Initialize Dinkelbach parameter $\lambda^{(0)} = \frac{\mathcal{N}(\mathbf{w}^{(0)}, \boldsymbol{\alpha}^{(0)})}{\mathcal{D}(\mathbf{w}^{(0)}, \boldsymbol{\alpha}^{(0)})}$.\\
Set outer iteration index $t = 0$.

\While{$|\mathcal{N}(\mathbf{w}^{(t)}, \boldsymbol{\alpha}^{(t)}) - \lambda^{(t)} \mathcal{D}(\mathbf{w}^{(t)}, \boldsymbol{\alpha}^{(t)})| > \epsilon_{out}$}{
    
    Set inner iteration index $i = 0$, $j = 0$, $\mathbf{w}^{(t,0)} = \mathbf{w}^{(t)}$, $\boldsymbol{\alpha}^{(t,0)} = \boldsymbol{\alpha}^{(t)}$.

    \While{$\|\boldsymbol{\alpha}^{(t,i+1)} - \boldsymbol{\alpha}^{(t,i)}\|_F^2 > \epsilon_1$ and $\|\mathbf{W}^{(t,i+1)} - \mathbf{W}^{(t,i)}\| > \epsilon_2$}{
        Transform (P3) to an SDP problem through  \eqref{eq:obj_W}\eqref{eq:rate_convex_W}\eqref{eq:sens_W}.\\
        Solve the convex SDP in (P5) \eqref{P5} via CVX to obtain $\mathbf{W}^*$.\\

    Relax discrete constraint $\alpha_k \in \mathcal{A}_k$ to continuous $0 \le \alpha_k \le 1$.\\
        Apply SCA to linearize objective $\mathcal{F}_{\boldsymbol{\alpha}}$ and constraints using \eqref{eq:alpha_obj_sca}, \eqref{eq:rate_convex_alpha},  \eqref{eq:sensing_sca_linear}.\\
        Solve the convexified linear program P6 \eqref{eq:alpha_LP} to obtain continuous $\boldsymbol{\alpha}^{(t,i+1)}$.\\
       $i \leftarrow i + 1$
    }

    Update $(\mathbf{w}^{(t+1)}, \boldsymbol{\alpha}^{(t+1)}) \leftarrow (\mathbf{w}^{(t,i)}, \boldsymbol{\alpha}^{(t,j)})$.\\
    Update auxiliary variable $\lambda^{(t+1)}$ using \eqref{30}.\\
    $t \leftarrow t + 1$.
}
\ForEach{AIoT device $k=1$ to $K$}{
    Map continuous $\alpha_k^*$ to discrete set $\alpha_k^{\text{final}} = \arg\min_{\alpha \in \mathcal{A}_k} | \alpha - \alpha_k |$.
}

\Return{$\mathbf{w}^* = \mathbf{w}^{\text{final}}$, $\boldsymbol{\alpha}^* = \boldsymbol{\alpha}^{\text{final}}$.}
\end{algorithm}
\subsection{Algorithm Complexity Analysis}

The generalized Dinkelbach-based BCD algorithm is presented in Algorithm~\ref{alg:ISL_Maximization}, to solve the joint optimization problem (P1) by alternately updating the BS beamformer and the AIoT device modulations. The overall complexity primarily arises from the two subproblems within the inner BCD loop and the outer Dinkelbach iterations \cite{dinkelbach1967nonlinear,tseng2001convergence}. For the beamforming vector optimization, solving the SDP problem (P3) involves optimizing a matrix variable $\mathbf{W} \in \mathbb{C}^{M_t \times M_t}$. With $M_t$ transmit antennas and $N$ subcarriers, the complexity of solving this SDP problem is approximately $\mathcal{O}(\sqrt{M_t}(N M_t^4 + N^2 M_t^2 + N^3))$ \cite{boyd2004convex} using standard interior-point methods. For the convex programming reflection coefficient optimization, with $K$ AIoT devices and constraints governed by $N$ subcarriers, the complexity per iteration is dominated by $\mathcal{O}\left( (K + N)^3 \right)$. Let $T_{\text{out}}$ \cite{boyd2004convex} denote the number of outer Dinkelbach iterations, and $T_{\text{in}}$ denote the average number of inner BCD iterations. The quantization and feasibility check steps involve a single execution of the beamforming subproblem and a low-complexity projection, which are negligible compared to the iterative process. Therefore, the total computational complexity is given by 
\begin{equation}\label{eq:overall_complexity}
\mathcal{C}_t= \mathcal{O} (T_{\text{out}} T_{\text{in}}  (\sqrt{M_t}(N M_t^4 + N^2 M_t^2 + N^3)+ \
(K + N)^3)).
\end{equation}

Given that the number of antennas $M_t$ and AIoT devices $K$ are generally within a moderate range in practical scenarios, and considering the polynomial complexity scaling, the proposed algorithm can be efficiently executed by modern BS processors, thereby validating its practicality for real-world ISAC deployment \cite{peng2016recent,liu2022integrated}.

\begin{figure}[tp]
    \centering
    \includegraphics[width=\linewidth]{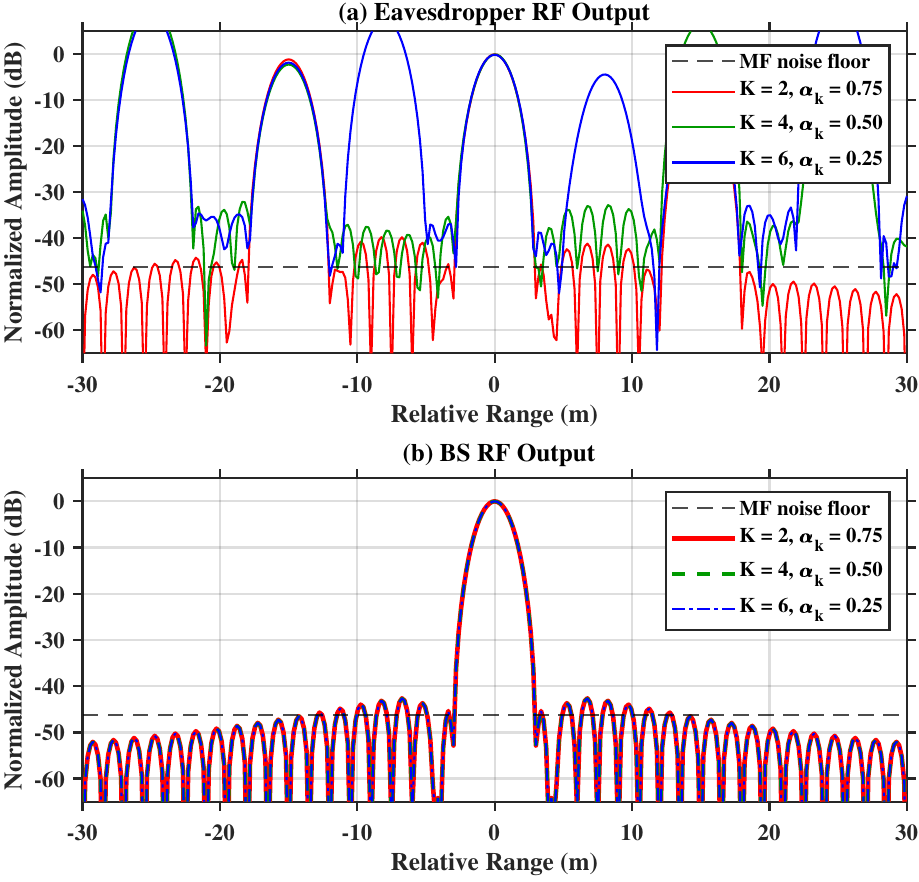}
    \caption{Range profiles for (a) SEve and (b) BS under the ISAC signals with the secure ACF, using the match filter.}
    \label{range}
\end{figure}

\section{Performance Evaluation} \label{performance}

In this section, we present numerical simulation results to validate AmbSentry's effectiveness. Unless stated otherwise, BS transmits an OFDM signal with  $N=128$ subcarriers operating over a 10~MHz bandwidth at a carrier frequency of 3~GHz \cite{11202391,11197530}. BS is equipped with uniform linear arrays \cite{11104928} consisting of $M_t=8$ transmit, and $M_r=8$ receive antennas, operating with a maximum transmit power of 30~dBm. The network topology places the communication user at a distance of 100~$m$ from the BS, while the sensing target is located at a distance of 30~$m$ from the BS \cite{11037613, liu2022integrated}. $\sigma_t$ is set to 2, representing a medium-sized target \cite{liu2022integrated}, such as a bicycle or a small vehicle. We evaluate the system's performance by varying the number of AIoT devices, $K$, from 2 to 10, with devices randomly distributed around the target. The proposed optimization framework maximizes the ISL using a Dinkelbach-based iterative algorithm, subject to a fixed sensing SINR constraint of 5~dB and varying communication rate constraint of 8~bps/Hz \cite{11202391}. Thermal noise floor is set as -174~dBm/Hz \cite{sengijpta1995fundamentals}. The random communication data is modulated from a 16~QAM constellation, where $\mu_2=1$ and $\mu_4=1.32$ \cite{proakis2001digital}.

\subsection{Validating Artificial Target Generation for Secure ACF Design}

The environment includes $K \in \{2, 4, 6\}$ AIoT devices distributed around the target, with angles ranging from $-20^\circ$ to $-60^\circ$, to simulate rich multipath scattering. For range profile visualization, an FFT size of 1024 is utilized. Fig.~\ref{range} presents the range profiles obtained by the MF at SEve and the BS, illustrating the effectiveness of AmbSentry. As shown in Fig.~\ref{range}(a), for SEve, the optimized reflection coefficients of the AIoT devices generate significant artificial clutter and high-level sidelobes. These sidelobes introduced by AIoT devices could effectively mask the mainlobe of the true target, making it difficult to distinguish the true target from the sidelobes. In contrast, Fig.~\ref{range}(b) demonstrates that the BS observes a clean range profile with a distinct target peak and significantly reduced sidelobes, confirming that the BS can effectively eliminate the sidelobe introduced by the AIoT devices by possessing knowledge of the CSI and the AIoT devices.

\subsection{Security and S\&C Trade-offs}



\begin{figure}[tp]
    \centering
    \includegraphics[width=\linewidth]{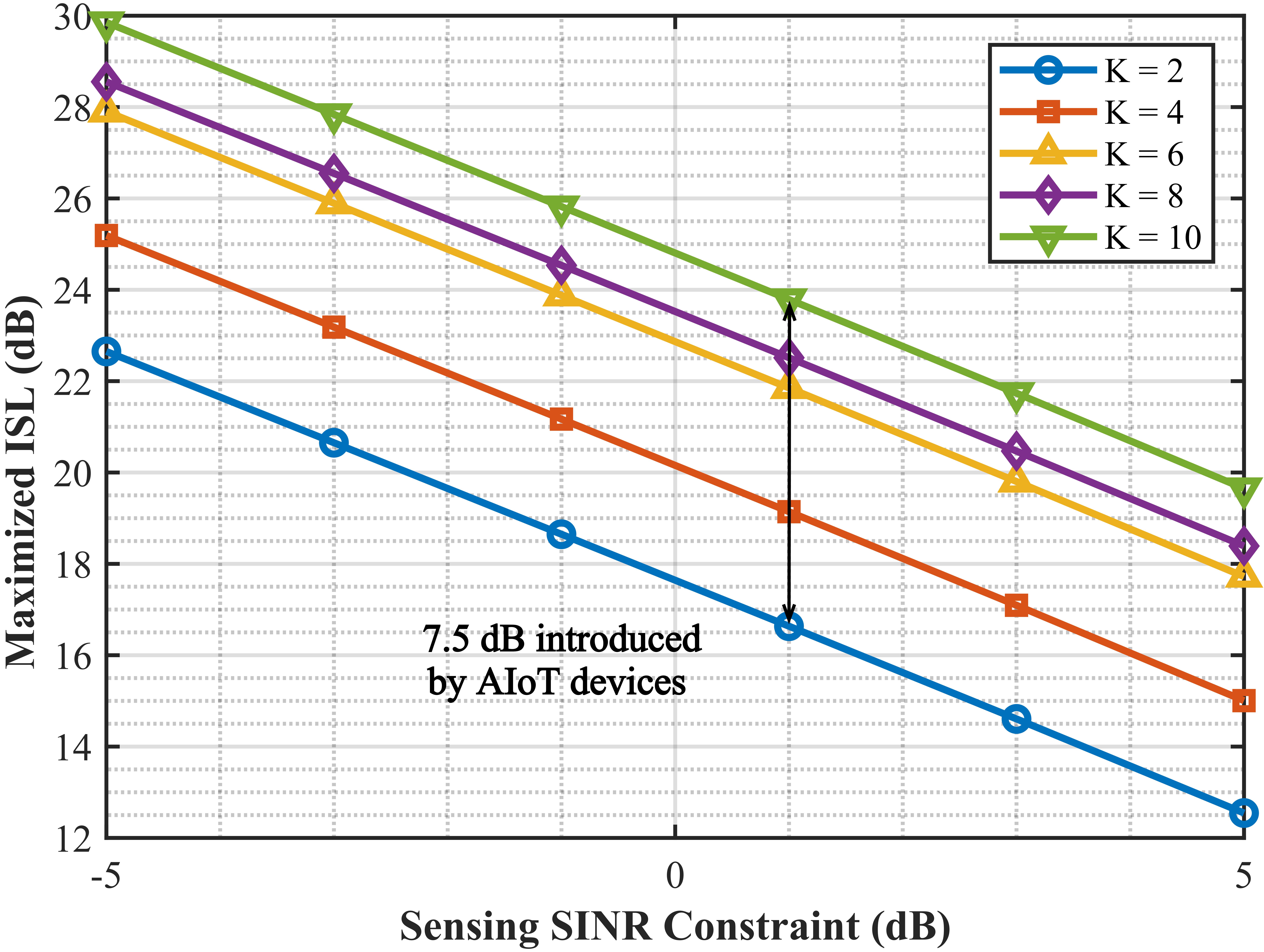}
    \caption{Trade-off between sensing SINR and sensing security under various numbers of AIoT devices.}
    \label{6}
\end{figure}

\begin{figure}[tp]
    \centering
    \includegraphics[width=\linewidth]{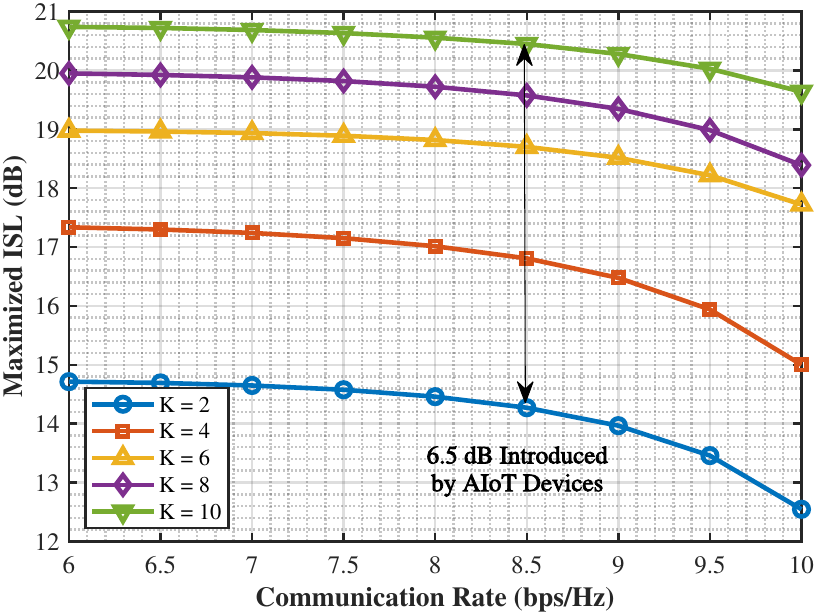}
    \caption{Trade-off between sensing security and communication rate under various numbers of AIoT devices.}
    \label{5}
\end{figure}

For graphical clarity, we set $M_t=M_r=16$. Fig.~\ref{6} investigates the impact of the sensing SINR on the maximized ISL under varying numbers of AIoT devices.
First, a monotonic decrease in the ISL is observed as the required sensing SINR increases. This trend highlights a direct conflict between sensing security and sensing performance. To satisfy a higher sensing SINR, the transmit beamformer must concentrate more energy into the mainlobe directed at the target. This concentration reduces the energy of the sidelobes to confuse SEve, thereby degrading the security performance against SEve.
Second, the results further validate the effectiveness of the AIoT device-aided ISAC system. Even under the strict sensing SINR constraint, increasing the number of AIoT devices from $K=2$ to $K=10$ provides a substantial performance gain (approximately 7.5 dB). This indicates that the additional passive paths provided by the AIoT devices can compensate for the loss of the DoFs, allowing the system to maintain a secure waveform even when high-precision sensing is demanded.

Fig.~\ref{5} illustrates the trade-off between the communication rate and the maximized ISL under varying numbers of AIoT devices. It is found that the ISL decreases monotonically as the communication rate increases. This phenomenon demonstrates a fundamental conflict between communication and sensing security: satisfying a stricter rate constraint consumes the available DoF in the transmit beamforming matrix and the AIoT device modulations, leaving less spatial flexibility to shape the waveform for optimal sidelobe interference at the SEve.
In addition, increasing the number of AIoT devices from $K=2$ to $K=10$ significantly improves sensing security performance under all rate constraints. This indicates that the AIoT devices introduce additional controllable multipath components, which the system effectively leverages to randomize the channel seen by SEve or enhance the correlation properties at the BS, thereby compensating for the performance loss caused by high communication demands. However, it is notable that the rate of improvement in ISL decreases as the number of AIoT devices increases. This saturation occurs because the transmit beamforming power is finite. Therefore, simply increasing the number of AIoT devices cannot improve the DoFs indefinitely without a corresponding increase in power resources.

\begin{figure}[tp]
    \centering
    \includegraphics[width=\linewidth]{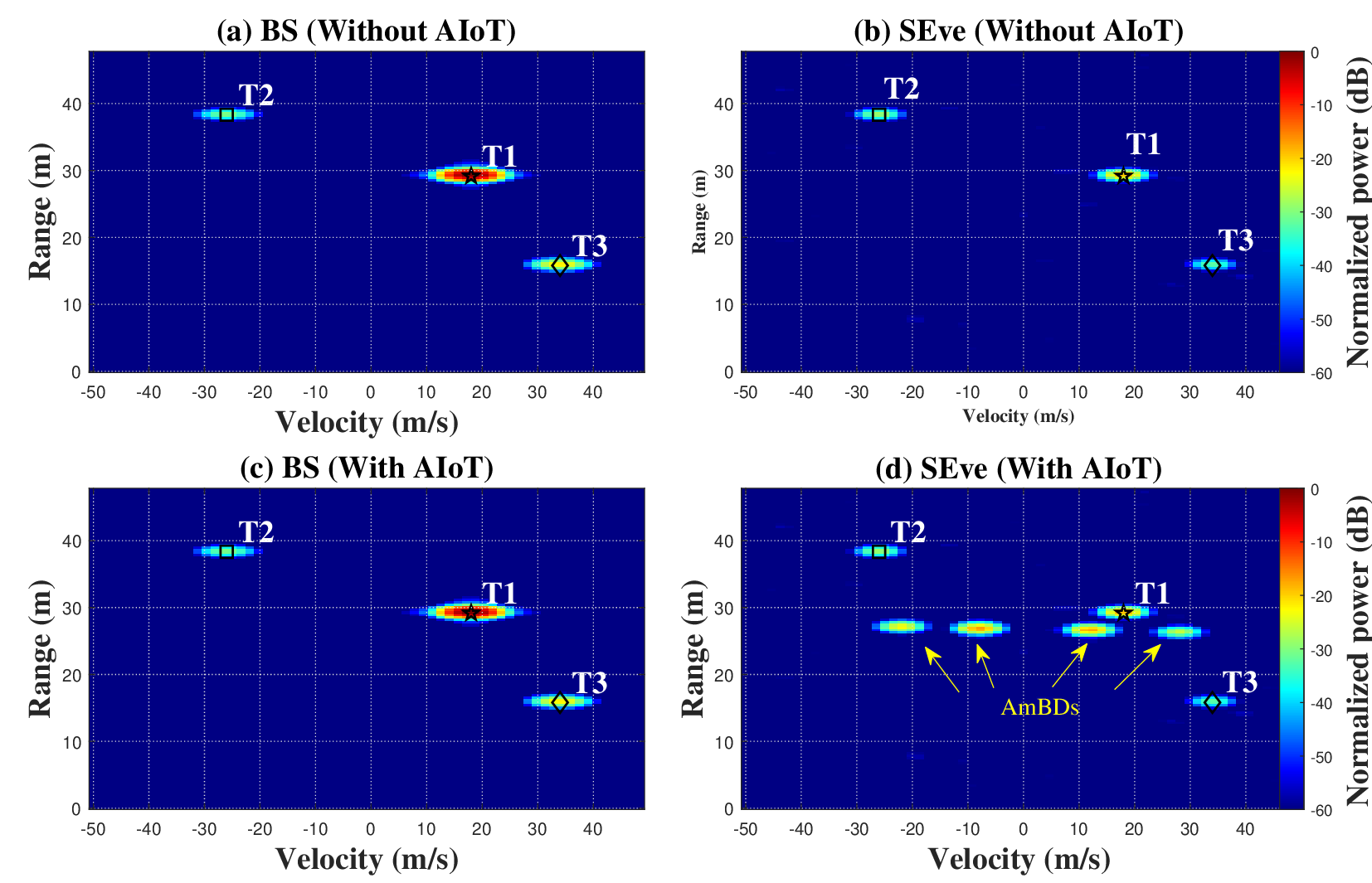}
    \caption{Range-Doppler maps for (a)(c) BS side and (b)(d) SEve side.}
    \label{7}
\end{figure}

\begin{figure}[tp]
    \centering
    \includegraphics[width=\linewidth]{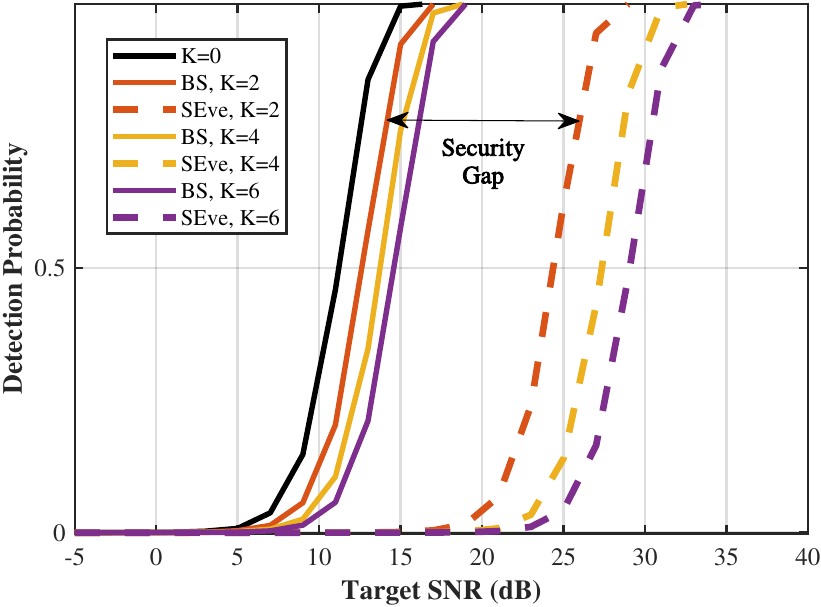}
    \caption{ Comparison of the target detection probability ($P_d$) between the BS and Eavesdropper versus SNR.}
    \label{8}
\end{figure}

\subsection{Security Performance Analysis: Detection Gap}
To validate the effectiveness of AmbSentry in the range-Doppler domain, Fig.~\ref {7} presents the range-Doppler  response maps with and without the assistance of AIoT devices for the BS and SEve. Comparing Fig.~\ref{7}(a) with Fig.~\ref{7}(c), it is directly seen that the introduction of AIoT devices does not negatively affect RD maps of the BS. Since the BS has prior knowledge of the AIoT device reflection coefficients and relevant CSI, it can effectively resolve target echoes from the superimposed signals, maintaining a clear and sharp target peak on the RD profile. Furthermore, the comparison between Fig.~\ref{7}(b) and Fig.~\ref{7}(d) demonstrates that it is difficult for SEve to mitigate the interference caused by the AIoT devices. Lacking the necessary prior information, SEve fails to eliminate the random modulated interference generated by the AIoT devices, resulting in an RD map dominated by high-power clutter. Most crucially, the distinct contrast between Fig.~\ref{7}(c) and Fig.~\ref{7}(d) indicates the sensing performance gap established between the BS and SEve. While the BS achieves a high SINR, SEve suffers from severe sidelobe interference and false detection of the artificial targets.

Fig.~\ref{8} directly evaluates the sensing security by comparing the target detection probability of various BS and SEve SNR values, given a fixed false alarm probability ($10^{-6}$).
As clearly illustrated, the BS achieves superior detection performance, with its detection probability rapidly approaching $1$ before the SNR  of 20 dB, where the detection probability of SEve approaches zero. Specifically, SEve requires approximately 14 dB higher SNR than the BS to achieve the same detection probability.
This validates the effectiveness of AmbSentry in maintaining the sensing performance at the legitimate sensing receiver and degrading the sensing performance of SEve. 
Consequently, the substantial performance gap between the two curves validates that AmbSentry successfully creates a sensing security zone, making the target hard to detect by SEve while maintaining high-reliability sensing for the BS.

\begin{figure}[tp]
    \centering
    \includegraphics[width=\linewidth]{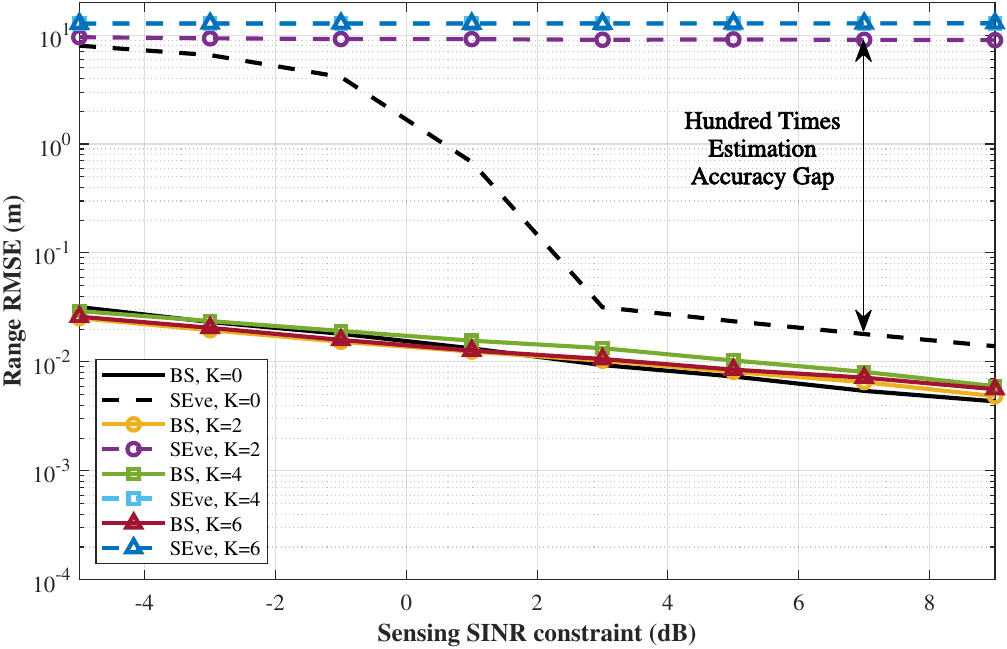}
    \caption{  Range estimation performance of BS and SEve, for various Sensing SINR constraints }
    \label{10}
\end{figure}

\subsection{Security Performance Analysis: Estimation Gap}

To further evaluate the performance of the practical estimation of the target parameters and the estimation gap between the BS and SEve, a range estimation simulation is conducted in
Fig.~\ref{10}, which is quantified by the root mean square error (RMSE), for both the BS and SEve under varying sensing-SINR constraints. Utilizing the root MUSIC estimator with a given source number~\cite{1143830}, the results demonstrate that legitimate sensing performance remains robust, despite residual clutter arising from the imperfect cancellation of AIoT device signals at the BS. In contrast, SEve’s estimation of RMSE persists at a high level regardless of SINR variations. This confirms the efficacy of AIoT devices as cooperative jammers. By introducing dynamic multiplicative noise, these AIoT devices effectively obfuscate target range parameters, ensuring a significant performance disparity between the legitimate receiver and the eavesdropper.

\begin{figure}[tp]
    \centering
    \includegraphics[width=\linewidth]{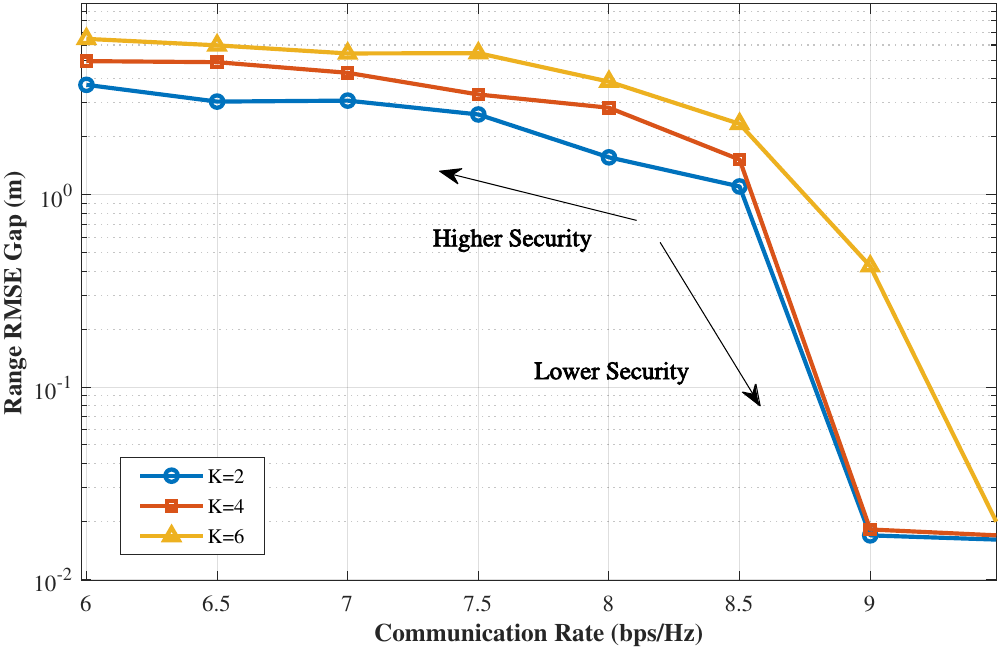}
    \caption{  Range estimation performance of BS and Eve, for various data rate constraints. The sensing SINR is set to 5 dB and $\eta$ = 0.05.}
    \label{11}
\end{figure}

Fig.~\ref{11} shows the trade-off between communication rate and sensing security, where we define the range RMSE gap (SEve’s RMSE - BS’s RMSE) between BS and SEve as a sensing secrecy metric. This gap is primarily attributed to the controllable artificial clutter introduced by the AIoT devices, which act as cooperative jammers to obfuscate unauthorized sensing. The results confirm that secure sensing functionality can be flexibly achieved at the expense of communication performance. It should be noted that introducing more AIoT devices further improves the security under a given communication rate constraint. This is because the increased number of AIoT devices could improve ISL, as shown in Fig. \ref{5}, which improves security.

\begin{figure}[tp]
    \centering
    \includegraphics[width=\linewidth]{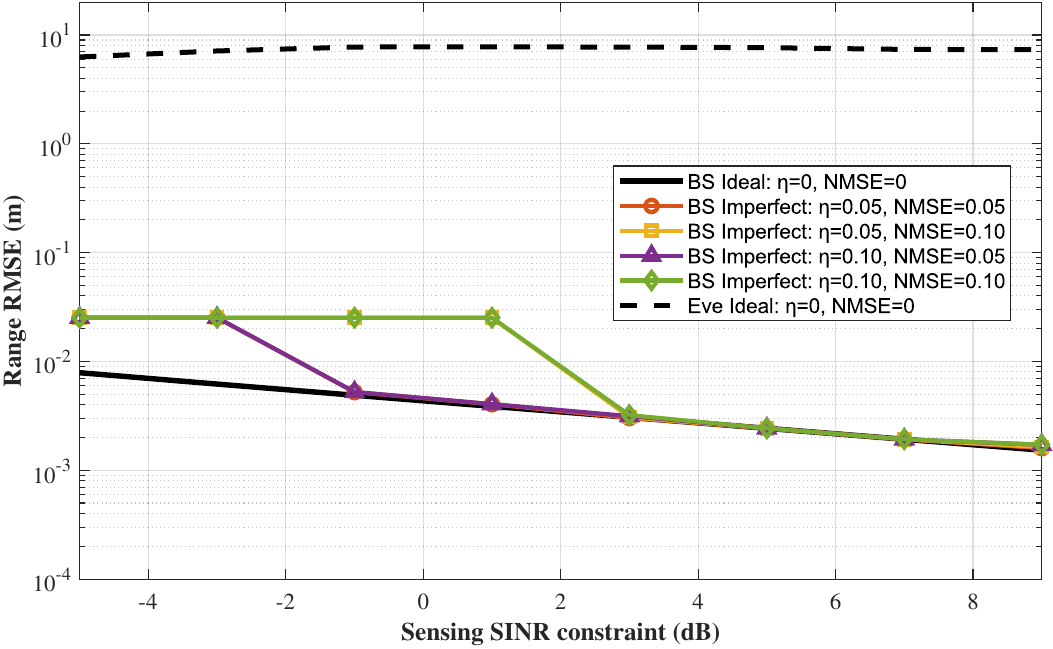}
    \caption{  Range estimation performance of BS under imperfect AIoT device cancellation and sensing-channel estimation errors. }
    \label{error}
\end{figure}

\subsection{Robustness Analysis: Imperfect Channel Estimation and AIoT device Signal Cancellation}
Fig.~\ref{error} illustrates the range estimation performance at the BS in the presence of imperfect AIoT device cancellation and sensing-channel estimation errors. 
The residual interference of the AIoT device after cancellation, quantified by the coefficient~$\eta$, effectively increases the level of disturbance in the BS, whereas the NMSE of the sensing channel results in a mismatch between the true composite sensing channel and the knowledge of the channel used in the joint beamforming and optimization of the AIoT device coefficient. 
Consequently, the optimized design no longer matches the true sensing condition, leading to a non-negligible increase in RMSE compared to the ideal case, especially when the sensing-SINR requirement is low.
Once the SNR exceeds a threshold (e.g., 0 dB), the RMSE of the BS distance estimation rapidly decreases and stabilizes at a low level, with minimal deviation from ideal performance. This shows that the proposed scheme exhibits strong robustness to imperfect interference. Specifically, as long as the SNR is sufficiently high, the residual AIoT device noise does not excessively degrade the base station's normal target perception performance. This validates the system's effectiveness under real-world imperfect conditions.

\section{Conclusion} \label{conclusion}

In this paper, we have proposed AmbSentry, an ISAC system that enhances sensing security by leveraging AIoT devices as passive, energy-efficient, and friendly jammers to impair unauthorized sensing. By explicitly characterizing the asymmetric impact of interference introduced by AIoT devices on eavesdroppers, we have demonstrated that sensing security can be fundamentally enhanced without relying on active AN injection, additional hardware deployment, or prior knowledge of the eavesdropper’s CSI. We have jointly optimized the active beamforming at the BS and the passive reflection coefficients of the AIoT devices to generate dynamic clutter, effectively maximizing the sensing security while guaranteeing QoS for legitimate S\&C receivers. The simulation results have demonstrated the effectiveness in degrading Eve’s target detection and estimation capabilities while maintaining robust legitimate sensing performance.
Specifically, the proposed system can achieve a detection-probability gap of approximately 14 dB and a two orders of magnitude estimation-error gap between the legitimate sensing receiver and the eavesdropper.
Furthermore, the findings have highlighted the inherent trade-offs in sensing-secure ISAC systems and provided valuable insights for secure system design. The proposed fundamental framework can be broadly extended to various research areas in ISAC, including secure sensing in highly dynamic environments with fast-time-varying channels, estimation- and information-theory-based secure ISAC signaling design, and joint constellation and coding optimization, by offering additional DoF to further enhance sensing security through deployed AIoT devices.

\bibliographystyle{IEEEtran}
\bibliography{IEEEabrv,reference}

\end{document}